\documentstyle[11pt]{article}
\input amssym.def
\input amssym.tex

\def\goth{\frak}
\def\double{\Bbb}

\def\cc{{\double C}}

\def\rr{{\double R}}

\def\qq{{\double Q}}
\def\mm{{{\cal M}}}

\def\ot{\otimes}
\def\op{\oplus}

\def\bb{\begin{eqnarray}}
\def\ee{\end{eqnarray}}
\def\eee{\nonumber\end{eqnarray}}

\newcommand{\tr}{\mathrm{tr}}
\newcommand{\alp}{\alpha}
\newcommand{\bet}{\beta}
\newcommand{\1}{\mathbf{1}}
\newcommand{\lam}{\lambda}
\newcommand{\diag}{\mathrm{diag}}
\newcommand{\Lam}{\Lambda}
\newcommand{\vphi}{\varphi}
\newcommand{\ten}{\otimes}
\newcommand{\Del}{\Delta}
\newcommand{\ol}{\overline}
\newcommand{\ul}{\underline}
\newcommand{\betl}{\beta_{\lambda}}
\newcommand{\arcoth}{\mathrm{arcoth}}
\newcommand{\na}{\nabla}
\newcommand{\tphi}{\tilde{\phi}}
\newcommand{\vp}{\varphi}
\newcommand{\bp}{\bar{\varphi}}

\begin{document}

%\hsize 15truecm
%\vsize 20truecm
\font\twelve=cmbx10 at 13pt
\font\eightrm=cmr8
\def\petit{\def\rm{\fam0\eightrm}}
\baselineskip 15pt

\begin{titlepage}
\title{Dirac Type Gauge Theories and the Mass of the Higgs Boson}
\author{J\"urgen Tolksdorf\thanks{email: juergen.tolksdorf@mis.mpg.de}\\
Max-Planck Inst. for Mathematics in the Sciences\\Leipzig, Germany\and
Torsten Thumst\"adter\thanks{tthumstaedter@tuev-nord.de}\\
University of Mannheim, Germany\thanks{Since Nov. 05 at the
T{\"U}V NORD EnSys, Dep. ETB, Hannover, Germany}}
\date{May 18, 2007}
\maketitle

\begin{abstract}
We discuss the mass of the (physical component of the) Higgs boson
in one-loop and top-quark mass approximation. For this the minimal
Standard Model is regarded as a specific (parameterized) gauge
theory of Dirac type. It is shown that the latter formulation, in
contrast to the usual description of the Standard Model, gives a
definite value for the Higgs mass. The predicted value for the
Higgs mass depends on the value addressed to the top mass
${\rm m}_{\mbox{\tiny T}}.$ We obtain
${\rm m}_{\mbox{\tiny H}} = 186 \pm 8 \;{\rm GeV}$ for
${\rm m}_{\mbox{\tiny T}} = 174 \pm 3\;{\rm GeV}$ (direct
observation of top events), resp.
${\rm m}_{\mbox{\tiny H}} = 184 \pm 22 \;{\rm GeV}$ for
${\rm m}_{\mbox{\tiny T}} = 172 \pm 10 \;{\rm GeV}$ (Standard
Model electroweak fit). Although
the Higgs mass is predicted to be near the upper bound,
${\rm m}_{\mbox{\tiny H}}$ is in full accordance with the range
$114 \leq {\rm m}_{\mbox{\tiny H}} < 193 \;{\rm GeV}$ that is allowed
by the Standard Model.\\

We show that the inclusion of (Dirac) massive neutrinos does not
alter the results presented. We also briefly discuss how the
derived mass values are related to those obtained within the frame
of non-commutative geometry.
\end{abstract}

\vspace{5cm}


\vspace{0.5cm}


\end{titlepage}

\section{Introduction}
The Higgs boson (more precisely, the physical component thereof)
is known to be the last outstanding particle predicted by the
(minimal) Standard Model (STM). Within the usual description of
the STM the expected mass range of the Higgs boson is restricted
to the interval $[114, 193) \;{\rm GeV}.$ (c.f. \cite{Ro:02}).
This prediction of the range of the Higgs mass results from
including quantum corrections and additional experimental input.
Of course, over the last decade there have been many attempts to
better specify the value of this mass range using different
mathematical approaches to the STM. One particular mathematical
approach worth mentioning here is given within the realm of
non-commutative geometry, see for instance \cite{Coq:89},
\cite{Kas:91}, \cite{Kas:92}, \cite{Kas:93}, \cite{GV:93},
\cite{Con:94}, \cite{KS:96} and more recently \cite{ChCoMa:06}.
For similar approaches one may consult, for instance,
\cite{HPS:91}, \cite{MO:94}, \cite{MO:96} and the
appropriate references therein.\\

In this paper we discuss the Higgs mass within the STM using the
geometrical frame of (parameterized) Dirac type gauge theories (GTDT).
The mathematical background of GTDT is discussed in some detail in
\cite{ToTh:05} and \cite{Tol:07}. However, in order to be self-contained
we present a purely local description of GTDT which is also needed to
derive the relations for the Higgs mass.\\

The basic idea of GTDT is to introduce a general geometrical setup
to describe (a certain class of) gauge theories in terms of
fermions. Hence, the fundamental ingredients of GTDT are so-called
``generalized Dirac operators''. They basically differ from the
usual Dirac operator by a general zero order term. This zero order
term in turn may be used to define a gauge potential not only by a
single one-form but also by forms of various degrees (please, see
below). In this sense, Dirac type operators may be regarded as more
general than connections. Physically speaking, generalized Dirac
operators permit incorporation of different fields into one
single mathematical object, which in turn are physically motivated
by the postulated interactions of the fermions considered. Another
advantage of describing gauge theories in terms of generalized Dirac
operators is that the latter naturally induce specific Lagrangian
densities. These densities can be shown to be equivariant with respect
to the action of the full gauge group including the gauge group of
Yang-Mills, of gravity and the diffeomorphism group of the
underlying (space-time) manifold. In this sense one may say that
the gauge theories defined by the corresponding Lagrangians have a
``square root'' in terms of generalized Dirac operators. This is
not only conceptually more satisfying than the usual ``adding of
actions'', but may also have some phenomenological consequences.
Accordingly, the present paper aims at showing how the geometrical
setup of GTDT allows the specification of the range of the Higgs
mass of the STM. The calculations presented are similar to those
given, for instance, in \cite{KaSch:97}. In particular, we restrict
our discussion to one-loop order and top-quark mass approximation.\\

One basic feature of GTDT is that it is logically inconsistent to
assume that space-time is flat. This is because a Dirac type operator
generically yields a non-vanishing energy-momentum tensor which in
turn implies a non-vanishing curvature of space-time. However, one
may still assume that gravitational effects may be negligible in
comparison with some given energy scale naturally implied by the
gauge theory at hand. In fact, this is our basic assumption as far
as the presented calculations of the Higgs mass are concerned (for
a corresponding justification see also the concluding remarks related
to this issue). Moreover, it is assumed that the Standard Model
(as well as perturbation theory) is valid up to a certain energy
scale $E_{\mbox{\tiny c}},$ which is much smaller than the Planck
scale but significantly higher than the scale set by the yet to find
Higgs mass. \\

The paper is organized as follows: In the second section we
present a purely local description of GTDT as it is needed to
follow the line of reasoning involved in the calculation of the
Higgs mass. In the third section we summarize the STM as it is
described as a special GTDT. There we also present a natural
parametrization of the general mathematical scheme that is
presented in \cite{ToTh:05}. In the fourth section we discuss
the parameter relations between the appropriately parameterized
GTDT of the STM with its usual mathematical description. We then
discuss the resulting renormalization flow equations for the
energy dependence of the coupling constants to one-loop order
and in top-quark mass approximation. This is done in the
$\overline{MS}-$scheme. Afterward we discuss the possible changes
when a massive neutrino sector is included. We also discuss in
this section the principal bounds of the Higgs mass within GTDT.
In the fifth section we compare our results with those presented
within the geometrical scheme of non-commutative geometry. We
conclude with some comments on the results discussed in this paper.
In an appendix we briefly summarize the relations between the
gauge couplings and the empirical parameters used in this paper.

\section{GTDT - A Local Description}
In this section we present a local description of gauge theories
of Dirac type in the case of a four dimensional (parallelizable)
Lorentzian manifold. This description will then be applied to the
(minimal) Standard Model in the next section in order to obtain some
statements about the mass of the (physical component of the) Higgs
boson.\\

Basically, a GTDT is given by the following
{\it universal (Dirac-) Lagrangian}:
\bb
\label{basiclagrangian}
{\cal L}_{\rm D} &:=& ({\bar\psi}iD\psi +
V_{\rm D})\,\mbox{\small$\sqrt{-|g|}$}\,d^4x,\\[0.2cm]
\label{diracpotential}
V_{\rm D} &\equiv&
\mbox{\small$\frac{\rm N}{2}$}\,r_{\!\mbox{\tiny M}} +
{\rm tr}\!\left(\gamma^{\mu\nu}[\theta_{\mu},\theta_{\nu}]\right) +
\mbox{\small$\frac{1}{8}$}\,g_{\mu\nu}\,
{\rm tr}\!\left(\gamma^\sigma\,[\theta_{\sigma},\gamma^\mu]
\gamma^\lambda\,[\theta_{\lambda},\gamma^\nu]\right).
\ee

Here, $|g|\equiv\det(g_{\mu\nu})$ and $r_{\!\mbox{\tiny M}}$ denotes
the Ricci scalar curvature with respect to the Lorentz metric
$g_{\mu\nu}$ of signature $-2.$ The Dirac
matrices $\gamma^\mu\in{\bf M}_{\rm N}(\cc)$ fulfill the
Clifford relation $\{\gamma^\mu,\gamma^\nu\}\equiv\gamma^\mu\gamma^\nu+
\gamma^\nu\gamma^\mu = -2\,g^{\mu\nu}{\bf 1}_{\rm N}$ with $g^{\mu\nu}$
being the inverse of $g_{\mu\nu}.$ Also, we use the
common abbreviation $\gamma^{\mu\nu}\equiv
\mbox{\small$\frac{1}{2}$}\,[\gamma^\mu,\gamma^\nu]$ for the
generators of the (proper orthochroneous) Lorentz transformations
in the spin representation. The dimension ${\rm N}\equiv{\rm 4N}_{\rm F}$
of the representation space is given by the ``fermion representation'',
i.e. $\psi\in{\cal C}^\infty(\mm,\cc^4\ot\cc^{{\rm N}_{\rm F}}).$
In the following $\mm\subset\rr^4$ denotes some open (connected) subset
such that $T\mm \simeq \mm\times\rr^4.$\\

The Dirac operator $D = \gamma^\mu\nabla^{\mbox{\tiny D}}_{\!\!\mu}$ is defined in
terms of the (Dirac) connection
\bb
\label{diracconnection}
\nabla^{\mbox{\tiny D}}_{\!\!\mu} &:=& \partial_{\!\mu} + \omega_\mu + \theta_{\!\mu}\cr
&\equiv&
\nabla^{\mbox{\tiny S}}_{\!\!\mu} + \theta_{\!\mu},
\ee
with $\omega_\mu\equiv
\mbox{\small$\frac{1}{4}$}\gamma^{\alpha\beta}\omega_{\mu\alpha\beta}
\in{\cal C}^\infty(\mm,{\bf M}_{\rm N}(\cc))$ being the spin-connection form
and $\nabla^{\mbox{\tiny S}}_{\!\!\mu}$ the corresponding spin connection
with respect to $g_{\mu\nu}.$ Also, the one-form
$\theta_{\!\mu}\in{\cal C}^\infty(\mm,{\bf M}_{\rm N}(\cc))$
denotes a general gauge potential. The connection (\ref{diracconnection})
is called a {\it Clifford connection} (or, ``twisted spin-connection'') if
the general gauge potential $\theta_{\!\mu}$ fulfills
\bb
\label{cliffordrel}
[\theta_{\!\mu},\gamma^\nu] = 0.
\ee
In this case we write $\theta_{\!\mu} = A_\mu,$ such that a Clifford
connection reads\footnote{In \cite{ToTh:05} the covariant derivative of a
Clifford connection is denoted by $\partial_{\!A}.$ It should not be
confounded with the lifted Levi-Civita connection.}
\bb
\label{cliffordconnection}
\nabla^{\mbox{\tiny D}}_{\!\!\mu} &\equiv& \nabla_{\!\!\mu}^{\mbox{\tiny Cl}}
:=\; \partial_{\!\mu} + \omega_\mu + A_\mu\cr
&=&
\nabla^{\mbox{\tiny S}}_{\!\!\mu} + A_\mu.
\ee
Accordingly, the Dirac operator $D$ is then called a ``twisted spin Dirac
operator''. However, for a general gauge potential $\theta_{\!\mu}$ one has
$[\theta_{\!\mu},\gamma^\nu] \not= 0.$ In this more general situation the
appropriate Dirac operator $D$ is known as a {\it generalized Dirac operator}
(or, ``operator of Dirac type''), see for example in \cite{AtBoSh:64} and
\cite{BGV:91}. In what follows, however, we will refer to
$D=\gamma^\mu\nabla^{\mbox{\tiny D}}_{\!\!\mu}$
simply as a Dirac operator, even in the case where $D$ is defined with
respect to general $(g_{\mu\nu},\theta_{\!\mu}).$\\

Of course, a general gauge potential $\theta_{\!\mu}$ can be
decomposed as $\theta_{\!\mu} = A_\mu + (\theta_{\!\mu} - A_\mu)\equiv
A_\mu + H_\mu.$
It can be shown that the {\it Dirac potential} (\ref{diracpotential})
is independent of such a decomposition. Hence, without loss
of generality we may decompose a general connection (\ref{diracconnection})
into a Clifford connection plus a general gauge potential:
\bb
\nabla^{\mbox{\tiny D}}_{\!\!\mu} &=& \nabla_{\!\!\mu}^{\mbox{\tiny S}} + A_\mu +
H_\mu\cr
&=&
\nabla_{\!\!\mu}^{\mbox{\tiny Cl}} + H_\mu
\ee
and thereby substitute $\theta_{\!\mu}$ by $H_\mu$ in (\ref{diracpotential}).
This general gauge potential $H_\mu$ can be expressed also
in terms of the Dirac operator $D$ itself:
\bb
H_\mu &=& -\mbox{\small$\frac{1}{4}$}\,g_{\mu\nu}\gamma^\nu
(D-\gamma^\sigma\nabla_{\!\!\sigma}^{\mbox{\tiny Cl}})\cr
&\equiv&\xi_\mu\Phi_{\mbox{\tiny D}}.
\ee
Note that $\gamma^\mu\xi_\mu =
-\mbox{\small$\frac{1}{4}$}\,g_{\mu\nu}\gamma^\mu\gamma^\nu = {\bf 1}_{\rm N}$
and $\Phi_{\mbox{\tiny D}}=\gamma^\mu H_\mu.$ Thus, we may
decompose any (generalized) Dirac operator as
\bb
D = \gamma^\sigma\nabla_{\!\!\sigma}^{\mbox{\tiny Cl}} + \Phi_{\mbox{\tiny D}}.
\ee

Note that for any gauge potential $H'_\mu$ which fulfills the two
requirements: $\gamma^\mu H'_\mu = \gamma^\mu H_\mu$ and
$\xi_\mu\gamma^\nu H'_\nu = H'_\mu,$ one infers that
$H'_\mu = H_\mu.$ Also note that
\bb
\label{zeroorderdop}
\Phi_{\mbox{\tiny D}} = \sum_{k=0}^4\;\sum_{0\leq\nu_1<\cdots<\nu_k\leq 3}\,
\gamma^{\nu_1}\cdots\gamma^{\nu_k}\,\chi^{\mbox{\tiny$(k)$}}_{\nu_1\cdots\nu_k},
\ee
with
$[\chi^{\mbox{\tiny$(k)$}}_{\nu_1\cdots\nu_k},\gamma^\mu]=0$ being
considered as k-forms on $\mm$ which take values in ${\bf M}_{\rm
N}(\cc).$ The lowest order contribution $\chi^{(0)}$ is
characterized by $[\Phi_{\!\mbox{\tiny D}},\gamma^\mu]=0.$ In
contrast, the highest order contribution
\bb
\sum_{0\leq\nu_1<\cdots<\nu_4\leq 3}\,
\gamma^{\nu_1}\cdots\gamma^{\nu_4}\,\chi^{(4)}_{\nu_1\cdots\nu_4}
= \gamma_5\phi,
\ee
fulfills the condition
\bb
\label{highestorderrel} \{\Phi_{\!\mbox{\tiny D}},\gamma^\mu\}=0,
\ee
with $\phi\in{\cal C}^\infty(\mm,{\bf M}_{{\rm N}_{\rm F}}(\cc))$
and $\gamma_5 = i\gamma^0\cdots\gamma^3$ the canonical
grading operator on the spinor space (such that $\cc^4 =
S_{\rm L}\op S_{\rm R}$ decomposes into the ``left-handed'' and
``right-handed'' spinors). The condition (\ref{highestorderrel})
is analogous to (\ref{cliffordrel}) for it implies
\bb
[\theta_\mu,\gamma^\nu] =
{\mbox{\small$-\frac{1}{2}$}}\,\delta^\nu_\mu\gamma^\lambda\theta_\lambda.
\ee

Moreover, the first order contribution only yields a re-definition of the
Yang-Mills gauge potential $A_\mu.$ Hence, in the sequel we will omit the
first order part in $\Phi_{\mbox{\tiny D}}.$\\

The {\it relative curvature} of a Dirac type operator is defined as
\bb
F^{\mbox{\tiny$\theta$}}_{\!\mu\nu} &:=& \nabla^{\mbox{\tiny S}}_{\!\!\mu}\theta_{\!\nu} -
\nabla^{\mbox{\tiny S}}_{\!\!\nu}\theta_{\!\mu} + [\theta_{\!\mu},\theta_{\!\nu}]\cr
&=&
\partial_{\!\mu}\theta_{\!\nu} -
\partial_{\!\nu}\theta_{\!\mu} + [\theta_{\!\mu},\theta_{\!\nu}] +
[\omega_\mu,\theta_{\!\nu}] - [\omega_\nu,\theta_{\!\mu}].
\ee
It naturally decomposes as
\bb
\label{pauliterm}
F^{\mbox{\tiny$\theta$}}_{\!\mu\nu} &=&
F^{\mbox{\tiny A}}_{\!\mu\nu} + F^{\mbox{\tiny A,H}}_{\!\mu\nu}\cr
&=&
F^{\mbox{\tiny A}}_{\!\mu\nu} + F^{\mbox{\tiny H}}_{\!\mu\nu}
+ \kappa^{\mbox{\tiny A,H}}_{\mu\nu}.
\ee
Here, $\kappa^{\mbox{\tiny A,H}}_{\mu\nu}:=[A_\mu,H_\nu] - [A_\nu,H_\mu]$
abbreviates the ``interaction term'' between the gauge potentials
$A_\mu$ and $H_\mu.$ The curvature
\bb
F^{\mbox{\tiny A,H}}_{\!\mu\nu} &:=&
F^{\mbox{\tiny D}}_{\!\mu\nu} - F^{\mbox{\tiny Cl}}_{\!\mu\nu}\cr
&=&
\nabla_{\!\!\mu}^{\mbox{\tiny Cl}}H_\nu-
\nabla_{\!\!\nu}^{\mbox{\tiny Cl}}H_\mu + [H_\mu,H_\nu]
\ee
denotes the relative curvature of $H_\mu$ with respect to
$(\omega_\mu,A_\mu)$ and $F^{\mbox{\tiny D}}_{\!\mu\nu}$ is the
curvature with respect to the Dirac connection (\ref{diracconnection}).
In the case of $H_\mu=0$ (i.e. $\theta_{\!\mu}=A_\mu$)
the relative curvature is called the ``twisting curvature'' of $D.$ Since
$[\omega_\mu,A_\nu]=0,$ the twisting curvature
$F^{\mbox{\tiny A}}_{\!\mu\nu} = \nabla^{\mbox{\tiny S}}_{\!\!\mu}A_\nu -
\nabla^{\mbox{\tiny S}}_{\!\!\nu}A_\mu + [A_\mu,A_\nu]$ coincides with the
usual Yang-Mills field strength, provided Clifford connections
$\nabla_{\!\!\mu}^{\mbox{\tiny Cl}}$ are identified with Yang-Mills
connections $\nabla_{\!\!\mu}^{\mbox{\tiny A}}\equiv\partial_{\!\mu} + A_\mu.$\\

There is a distinguished class of Dirac type operators called
{\it Dirac operators of Yukawa type} (c.f. \cite{ToTh:05}).
These operators are defined by
\bb
\Phi_{\mbox{\tiny D}} := \gamma_5\phi
\ee

Note that
$D$ is odd if and only if $\cc^{{\rm N}_{\rm F}} = E_{\rm L}\op E_{\rm R}$ and
$\phi$ is odd. Assuming that $\phi\not=0,$ it can be shown that the field equations
for Dirac operators of Yukawa type give rise to the existence of a constant (skew-hermitian)
matrix function ${\cal D}\in{\cal C}^\infty(\mm,{\bf M}_{{\rm N}_{\rm F}}(\cc))$
and a real-valued smooth function $h\in{\cal C}^\infty(\mm,\rr)$ such that
\bb
\phi = h{\cal D}.
\ee

This reduces the gauge symmetry group to the isotropy group of ${\cal D}.$
Accordingly, a Yukawa type Dirac operator is said to be in the
{\it unitary gauge} if it reads
\bb
D = \gamma^\sigma\nabla_{\!\!\sigma}^{\mbox{\tiny Cl}} + \gamma_5{\cal D}.
\ee

A Yukawa-type Dirac operator is said to represent a {\it fermionic
vacuum} if
\bb
\label{fermionicvacuum}
D &=& \gamma^\mu(\partial_{\!\mu} + \omega_\mu) + \gamma_5{\cal D}\cr
&\equiv& \gamma^\mu\nabla_{\!\!\mu}^{\mbox{\tiny S}} + \gamma_5{\cal D}
\ee
and $g_{\mu\nu}$ fulfills the Einstein equation
\bb
R_{\mu\nu} = \kappa_{\mbox{\tiny gr}}{\rm tr}{\cal D}^2\,g_{\mu\nu}\,.
\ee

A general Yukawa type operator is then considered as a perturbation of
a fermionic vacuum. Note that with respect to the latter any Yukawa type
operator corresponds to $(h,A_\mu,h_{\mu\nu}).$ Here, the metric
$h_{\mu\nu}$ is considered as a perturbation of $g_{\mu\nu}$ which
satisfies the Einstein equation with the energy-momentum tensor being
defined with respect to $(\psi,h,A_\mu).$ As already mentioned in the
introduction, we will neglect the influence of a non-flat space-time
and assume that $g_{\mu\nu} = h_{\mu\nu} \thickapprox \eta_{\mu\nu}.$
Some appropriate comments on a justification of this assumption will be
given in the conclusion.\\

To lowest order the metric $g_{\mu\nu}$ is fully determined by the spectrum
of ${\cal D}^2.$ In contrast, the field equations of a Yukawa type operator
do not determine either the Yang-Mills connection (i.e. the gauge potential
$A_\mu$), or the (physical component of the) Higgs field $h.$ For this
one has to slightly enlarge the class of Yukawa type operators,
which is referred to as {\it Pauli-Yukawa} type Dirac operators
(PDY). They are defined by Dirac operators of the form
\bb
\label{pdyop}
D &=&
\left(%
\begin{array}{cc}
  \gamma^\mu(\nabla_{\!\!\mu}^{\mbox{\tiny S}} + \theta_{\!\mu}) &
  -\mbox{\small$\frac{1}{2}$}\gamma^{\mu\nu}F^{\mbox{\tiny$\theta$}}_{\!\mu\nu} \\[0.1cm]
  \mbox{\small$\frac{1}{2}$}\gamma^{\mu\nu}F^ {\mbox{\tiny$\theta$}}_{\!\mu\nu} &
  \gamma^\mu(\nabla_{\!\!\mu}^{\mbox{\tiny S}} + \theta_{\!\mu}) \\
\end{array}%
\right)\nonumber\\[.3cm]
&\equiv&
\gamma^\mu(\nabla_{\!\!\mu}^{\mbox{\tiny S}} + \theta_{\!\mu})
+ {\cal
I}(\mbox{\small$\frac{1}{2}$}\gamma^{\mu\nu}F^{\mbox{\tiny$\theta$}}_{\!\mu\nu})
\ee
with the fermion representation space being doubled.
Here, the {\it Higgs gauge potential} reads $H_\mu := \xi_\mu\gamma_5\phi$ and
thus $\theta_{\!\mu} = A_\mu + \xi_\mu\gamma_5\phi.$ Moreover,
${\cal I}:=\mbox{\tiny$\left(%
\begin{array}{cc}
  0 & -1 \\
  1 & 0 \\
\end{array}%
\right)$}$ may be regarded as defining an additional complex structure for
${\cal I}^2 = -{\bf 1}_2.$ Hence, in the case of a twisted spin Dirac operator
(i.e. $\theta_{\!\mu} = A_\mu$) a PDY reduces to
$D=\gamma^\mu\nabla_{\!\!\mu}^{\mbox{\tiny Cl}}
+ {\cal I}(\mbox{\small$\frac{1}{2}$}\gamma^{\mu\nu}F^{\mbox{\tiny A}}_{\!\mu\nu}),$
where the second part formally looks like the well-known ``Pauli-term''.\\

The relative curvature $F^{\mbox{\tiny A,H}}_{\!\mu\nu}$ of the
Higgs gauge potential explicitly reads
\bb
\label{ymhrelcurvature}
F^{\mbox{\tiny A,H}}_{\!\mu\nu} =
\gamma_5(\xi_\mu[\nabla^{\mbox{\tiny A}}_{\!\!\nu},\phi] -
\xi_\nu[\nabla^{\mbox{\tiny A}}_{\!\!\mu},\phi]) + [\xi_\mu,\xi_\nu]\phi^2
\,.
\ee
Note that this (relative) curvature depends on the total field content
$(g_{\mu\nu},\phi,A_\mu).$\\

With respect to a fermionic
vacuum (\ref{fermionicvacuum}) one may consider Clifford connections which
are compatible with the vacuum, i.e. gauge potentials $A_\mu$ which satisfy
\bb
\label{gaugepotreducible}
[A_\mu,{\cal D}]=0\,.
\ee
In this case the interaction term $\kappa^{\mbox{\tiny A,H}}_{\mu\nu}$
vanishes identically and the relative curvature (\ref{pauliterm}) reduces to
\bb
F^{\mbox{\tiny$\theta$}}_{\!\mu\nu} =
F^{\mbox{\tiny A}}_{\!\mu\nu} + F^{\mbox{\tiny H}}_{\!\mu\nu}\,.
\ee
In fact, the latter relation is equivalent to the compatibility of
a Clifford connection with a fermionic vacuum.\\

The interaction term $\kappa^{\mbox{\tiny A,H}}_{\mu\nu}$ has a
simple physical meaning. With respect to the fermionic vacuum it
corresponds to the ``Yang-Mills mass matrix''. Indeed, the
``generalized Yang-Mills Lagrangian''
${\rm tr}F^{\mbox{\tiny$\theta$}}_{\!\mu\nu}F_{\!\mbox{\tiny$\theta$}}^{\!\mu\nu}$
clearly yields a term like
\bb \label{ymmassmatrix}
g^{\alpha\beta}g^{\mu\nu}{\rm tr}\kappa^{\mbox{\tiny A,H}}_{\mu\alpha}
\kappa^{\mbox{\tiny A,H}}_{\nu\beta}
&\sim&
(1 + h)^2 g^{\mu\nu}{\rm tr}({\cal D}\{A_\mu, A_\nu\}{\cal D})\cr
&\sim&
(1 + h)^2\,{\rm M}^2_{ab}\,g^{\mu\nu}A^a_\mu A^b_\nu,
\ee
with ${\rm M}^2_{ab}:={\rm tr}({\cal D}\{{\rm T}_a,{\rm T}_b\}{\cal D})$
being proportional to the (squared) Yang-Mills
mass matrix and $A_\mu = A^a_\mu{\rm T}_a$. Note that the
generators ${\rm T}_a\in{\rm\bf M}_{{\rm N}_{\rm F}}(\cc)$ refer
to the fermion representation. Moreover, ${\rm M}^2_{ab}$ equals
zero exactly for
those generators which commute with ${\cal D}.$\\

We emphasize that the ``generalized Pauli term''
$\gamma^{\mu\nu}F^{\mbox{\tiny$\theta$}}_{\!\mu\nu}$ in
(\ref{pdyop}) does not contribute to the fermionic part in
(\ref{basiclagrangian}) when restricted to the real sub-space
of ``particles-anti-particles''. It only contributes to the Dirac
potential $V_{\rm D}.$ More precisely, for $D$ of Pauli-Yukawa
type one obtains the total Lagrangian:
\bb
\label{stmlagrangian}
{\cal L}_{\rm D} &=&
(2{\bar\psi}(i\gamma^\mu\nabla_{\!\!\mu}^{\mbox{\tiny Cl}} +
i\gamma_5\phi)\psi +
V_{\rm D})\,\mbox{\small$\sqrt{-|g|}$}\,d^4x,\\[0.2cm]
\label{bosonstmlagrangian}
V_{\rm D} &=& \lambda_{\mbox{\tiny gr}}\,r_{\!\mbox{\tiny M}} -
(\lambda_{\mbox{\tiny YM}}\,{\rm tr}{F^{\mbox{\tiny A}}_{\!\mu\nu}}^{\!\dagger}
F_{\mbox{\tiny A}}^{\mu\nu} +
\lambda_{\mbox{\tiny H}}\,{\rm tr}\nabla_{\!\!\mu}\phi^\dagger\nabla^{\mu}\phi -
V_{\mbox{\tiny H}}).
\ee
Here, $\nabla_{\!\!\mu}\phi \equiv [\nabla^{\mbox{\tiny A}}_{\!\!\mu},\phi]
= \partial_{\!\mu}\phi + [A_\mu,\phi]$ and
\bb
V_{\!\mbox{\tiny H}} = \alpha_{\mbox{\tiny H}}({\rm tr}\phi^\dagger\phi)^2 -
\beta_{\mbox{\tiny H}}{\rm tr}\phi^\dagger\phi
\ee
is the usual {\it Higgs potential} of the (minimal) Standard Model and
$\lambda_{\mbox{\tiny gr}},\lambda_{\mbox{\tiny YM}},
\lambda_{\mbox{\tiny H}},\alpha_{\mbox{\tiny H}},\beta_{\mbox{\tiny H}}$
are real parameters to be specified in the next section. For a more detailed
discussion, in particular, of the occurrence of the factor 2 and of
the grading involution $\gamma_5$ in the Yukawa coupling and the geometrical
meaning of (\ref{ymmassmatrix}), we again refer to \cite{ToTh:05}.\\

In the next section we will make use of
(\ref{stmlagrangian}) -- (\ref{bosonstmlagrangian}) which formally looks like
the total Lagrangian of the (minimal) Standard Model including gravity.
In fact, for specific Yukawa type Dirac operators one may appropriately
re-write the Dirac-Lagrangian (\ref{stmlagrangian}) to get exactly the form
of the STM-Lagrangian. The scheme proposed may also have some phenomenological
consequences since (\ref{stmlagrangian}) is derived in one stroke
by a specific class of Pauli-Yukawa type operators. For this, however,
one still has to take into account the different mass dimensions of the
various fields and to also include an appropriate parametrization of both
the general Dirac-Lagrangian (\ref{basiclagrangian}) and the specific class of
Dirac operators one deals with. Of course, the parametrization cannot be
arbitrary. It has to be compatible with the geometrical setup. Basically,
the motivation of an appropriate parametrization of the geometrical scheme
considered comes from physics and is analogues to the introduction of the
gauge coupling constants\footnote{These should not being regarded as a
generalization of the electric charge, for the mathematical origin of the
latter is quite different from the gauge coupling constants.} in ordinary
(non-Abelian) Yang-Mills gauge theories (please, see below). Clearly,
the possible phenomenological implications of a specific geometrical
description, for example of the STM, strongly depend on the parametrization
of the geometrical scheme considered (c.f. our discussion in Sec. 5).

\section{The (minimal) STM as a GTDT}
As discussed in the previous section the STM-Lagrangian has a
natural ``square root'' in terms of Pauli-Yukawa type Dirac operators
(PDY). This holds true in particular for the Higgs sector.
However, within the framework of Dirac type gauge theories the
Higgs field $\phi$ transforms with respect to the full fermion representation
of the gauge group. This is in contrast to the minimal Standard Model where
the Higgs field is supposed to transform with respect to a specific
sub-representation of the fermion representation $\rho_{\rm F}$
(see below). These two representations are related by the
Yukawa-coupling matrix $G_{\rm Y}$ which can be considered as a
linear mapping from the representation space of the Higgs field to
the representation space of left-handed fermions. For a general
discussion we again refer to \cite{ToTh:05}. In what follows, we
will restrict ourselves to the specific case of the minimal
Standard Model (see, for instance, \cite{Na:86}).

\subsection{Data of the (minimal) STM as a specific GTDT}
To specify a GTDT one has to choose a gauge group G, a unitary
representation $\rho_{\rm F}$ thereof, as well as some (class of)
Dirac operators $D$. In the case of the (minimal) STM these data
are specified by (here, we adopt the same notation as it was used
in \cite{tol:98}):
\begin{itemize}
\item
${\rm G}$ equals ${\rm SU(3)}\times{\rm SU(2)}\times{\rm U(1)};$
\item
$\rho_{\rm F}$ equals the fermion representation:
\bb
\label{fermionrep1}
{\rm E}_{\rm L} &:=&
\bigoplus_1^3\left[(1, 2, -1/2)\op(3, 2, 1/6)\right],\nonumber\\[.5em]
{\rm E}_{\rm R} &:=&
\bigoplus_1^3\left[(1, 1, -1)\op(3, 1, -1/3)
\op(3, 1, 2/3)\right],
\ee
where $(n_3, n_2, n_1)$ denote the tensor product,
respectively, of an $n_3$ dimensional representation of SU(3),
an $n_2$ dimensional representation of SU(2) and a one
dimensional representation of U(1) with "hypercharge" $y$:
$\rho(e^{i\theta}):=e^{iy\theta},\;y\in\qq,\,
\theta\in[0,2\pi[$.\\

More explicitly, we have
\bb
\rho_{\rm F}:=\rho_{\rm L}\op\rho_{\rm R}\!:
{\rm SU(3)}\times{\rm SU(2)}\times{\rm U(1)}\rightarrow
{\rm Aut}(E)\subset{\rm U}(45) \label{frep}
\ee
with
\bb
\rho_{\rm L}({\bf c}, {\bf w},b) &:=&
\pmatrix{{\bf c}\ot{\bf 1}_{\rm N}\ot
{\bf w}\,b^q_{\rm L}  &  0\cr
                            \phantom{a}  &  \phantom{b}\cr
0  & {\bf 1}_{\rm N}\ot {\bf w}\,b^l_{\rm L}},\\[.5cm]
\label{B-matrix}
\rho_{\rm R}({\bf c}, {\bf w},b) &:=&
\pmatrix{{\bf c}\ot{\bf 1}_{\rm N}\ot
{\bf B}^q_{\rm R}  &  0\cr
                              \phantom{a}  &  \phantom{b}\cr
                                    0  &  {\bf B}^l_{\rm R}}
\ee
and
\bb
\label{fermionrep2}
E &\equiv& E_{\rm L}\op E_{\rm R}\nonumber\\[.2cm]
&\simeq&
\left[(\cc^{18}_q\op\cc^{6}_l)\right]_{\rm L}\op
\left[(\cc^{9}\op\cc^{9})\!_q\op
\cc^{\rm 3}_l\right]_{\rm R}.
\ee

\item
The Dirac operator $D$ is of Pauli-Yukawa type with Yukawa
coupling $\phi$ given by
\bb
\phi \equiv
i\left(%
\begin{array}{cc}
  0 & {\tilde\phi} \\
  {\tilde\phi}^{\,\dagger} & 0 \\
\end{array}%
\right)
\ee
with
\bb
\label{higgsfield}
{\tilde\phi}\equiv G_{\rm Y}(\varphi) &:=&
\pmatrix{{\bf 1}_3\ot({\bf g}'^q\ot\varphi,\,
{\bf g}^q\ot\hbox{\boldmath$\epsilon$\unboldmath}
                    \overline{\varphi} ) & 0\cr
                    \phantom{a}  &  \phantom{b}\cr
                            0  &  {{\bf g}^l\ot\varphi}}\nonumber\\[.5cm]
&\equiv&\pmatrix{{\bf 1}_3\ot{\tilde\varphi}_q & 0\cr
                                     0  &  {\tilde\varphi}_l}.
\ee

Here, respectively,
${\bf g}'^q, {\bf g}^q\!\in\!{\bf M}_{\rm N}(\cc)$ denote
the matrices of the Yukawa coupling constants for quarks
of electrical charge -1/3 and 2/3 (i.e. of quarks of "d"-type,
and of "u"-type) and ${\bf g}^l\in{\bf M}_{\rm N}(\cc)$
is the matrix of
the Yukawa coupling constants for the leptons of charge -1
(i.e. of leptons of "electron" type). While ${\bf g}^q$
and ${\bf g}^l$ can be assumed to be diagonal and real,
the matrix ${\bf g}'^q$ is related to the Kobayashi-Maskawa
matrix and therefore is neither diagonal nor real. The
"weak hyper-charges" for the left and right handed quarks
(indicated by the superscript "q") and leptons (superscript "l")
are defined by: $\rho(b):=e^{iy\theta},\, b\in{\rm U}(1),\,
y\in\qq,\, \theta\in[0,2\pi[.$ Then, the
two by two diagonal matrices ${\bf B}^q_{\rm R}$ and
${\bf B}^l_{\rm R}$ in (\ref{B-matrix}) are:
${\bf B}^q_{\rm R}:={\rm diag}(b_{\rm R}^{d'}, b_{\rm R}^{u})$
and ${\,\bf B}^l_{\rm R}:=b^l_{\rm R}{\bf 1}_{\rm N}$. Here,
$b^u_{\rm R} := e^{i y^u_{\rm R}\theta},$ with
$y^u_{\rm R}$ being the hypercharge of the right-handed quarks
of ``u-type'' and similar for the other quarks and leptons.\\

In (\ref{higgsfield}) $\varphi\in{\cal C}^\infty(\mm,\cc^2)$
denotes the complex Higgs field of the minimal Standard Model.
It carries the specific sub-representation $\rho_{\rm H}$ of the
fermion representation $\rho_{\rm F}:$
\bb
\rho_{\rm H}:\,{\rm SU}(3)\times{\rm SU}(2)\times {\rm U}(1)
&\longrightarrow&{\rm U}(2)\cr
({\bf c},{\bf w}, b)&\mapsto&{\bf w}\,e^{iy_h\theta}.
\ee

Finally, $\hbox{\boldmath$\epsilon$\unboldmath}$ is the
anti-diagonal matrix
$\hbox{\boldmath$\epsilon$\unboldmath}
:={\phantom{-}0\;1\choose-1\;0},$ which intertwines the fundamental
representation of SU(2) and its conjugate complex representation, and
$\overline{\varphi}$ here means the complex conjugate of
$\varphi$.\\

As a sub-representation of $\rho_{\rm F}$ the representation $\rho_{\rm H}$
is fixed by the relations of the hyper-charges of the quarks and the leptons:
\bb
\label{hyperchargerel}
y_h &=& y^l_{\rm L} - y^l_{\rm R}\cr
                 &=& y^q_{\rm L} - y^{d'}_{\rm R}\cr
                 &=& y^u_{\rm R} - y^q_{\rm L}\,.
\ee
Here,
\bb
 \label{hyperchargeval}
(y^q_{\rm L}, \,y^l_{\rm L}) &=& (1/6, -1/2),\nonumber\\[.1cm]
(\,(y^{d'}_{\rm R}, y^u_{\rm R}),\,y^l_{\rm R}) &=&
(\,(-1/3,2/3),\,-1)\,,
\ee
according to the fermion representation (\ref{fermionrep1}).\\
\end{itemize}

The corresponding fermionic vacuum ${\cal D}$ is given by
\bb
{\cal D} := i\pmatrix{       0       & {\bf M}\cr
                                       {\bf M}^\dagger& 0}\,,
\ee
with
\bb
{\bf M} &\equiv& \pmatrix{{\bf 1}_3\ot{\bf M}_q & 0\cr
     0 & {\bf M}_l},\nonumber\\[.5em]
{\bf M}_q &\equiv&\pmatrix{      0         & {\bf m}^{d'}\cr
{\bf m}^u  &         0},\nonumber\\[.5em]
{\bf M}_l &\equiv& {0 \choose {\bf m}^l},
\ee

where, respectively, the matrices
${\bf m}^l\!:=\!\frac{v}{\sqrt{2}}\,{\bf g}^l\!\in
\!{\bf{\rm M}}_{\rm N}(\cc)$ and
${\bf m}^u\!:=\!\frac{v}{\sqrt{2}}\,{\bf g}^q\!\in
\!{\bf{\rm M}}_{\rm N}(\cc)$
denote the "mass matrices" of the charged leptons (l) and
quarks (q) of ``u-type". They can be assumed
to be diagonal and real. The corresponding
${\rm N}\!\times\!{\rm N}$ matrix
${\bf m}^{d'}\!:=\!\frac{v}{\sqrt{2}}\,{\bf g}'^q$ of ``d-type"
quarks is neither diagonal nor real. It is related to the
mass matrix of ``d-type" quarks
${\bf m}^d=\hbox{diag}(m^{d_1},\dots, m^{d_{\rm N}}),\,
m^{d_k}\!\in\!\rr,$ via the
Kobayashi-Maskawa matrix ${\bf V}\!\in\!{\rm U}(\rm N)$:
${\bf m}^{d'}\!=\!{\bf V}\,{\bf m}^{d}\,{\bf V}^*$. Here and
above ${\rm N}$ is of appropriate size that is defined by the fermion
representation (\ref{fermionrep1}) (or (\ref{fermionrep2})) and $v > 0$
is the ``vacuum expectation value'' of the Higgs boson.\\

With the choice of $({\rm G},\rho_{\rm F},D)$ we have mathematically
specified a particular Dirac type gauge theory. However, from a
physical perspective we still have to appropriately parameterize
this GTDT. Of course, the parametrization cannot be arbitrary. It
should be in accordance with the geometrical frame of a GTDT. As
we mentioned already, there are basically two objects which can be
parameterized: the Dirac-Lagrangian (\ref{basiclagrangian}) and
the general Dirac operator defined by (\ref{diracconnection}).
Note that in the specific case at hand the parameters introduced by
the Yukawa coupling matrices only arise because of the change of the
representation of the Higgs field (i.e. to consider $\phi$ as a
function of $\varphi$).

\subsection{Geometrical Parametrization}
In general, an admissible parametrization of (\ref{basiclagrangian})
is given by the commutant of the fermion representation $\rho_{\rm F}.$
This is similar to the introduction of the Yang-Mills coupling
constant for each simple gauge group. To explain this let us
consider the case where ${\rm G}={\rm SU(2)},$ $\rho_{\rm F}$ is some
unitary representation thereof and $D$ is specified by (\ref{pdyop})
with $\theta_{\!\mu} = A_\mu.$ In this case, the Dirac-Lagrangian
(\ref{stmlagrangian}) reduces to
\bb
\label{ehymlagrangian}
{\cal L}_{\rm D} &=&
(2{\bar\psi}i\gamma^\mu\nabla_{\!\!\mu}^{\mbox{\tiny Cl}}\psi +
V_{\rm D})\,\mbox{\small$\sqrt{-|g|}$}\,d^4x,\\[0.2cm]
\label{ehympotential} V_{\rm D} &=& \lambda_{\mbox{\tiny gr}}\,r_{\!\mbox{\tiny M}}
- \lambda_{\mbox{\tiny YM}}\,{\rm tr}{F^{\mbox{\tiny A}}_{\!\mu\nu}}^{\!\!\dagger}
F_{\!\mbox{\tiny A}}^{\mu\nu}.
\ee

Here, the constants $\lambda_{\mbox{\tiny gr}}, \lambda_{\mbox{\tiny YM}}$
are purely numerical and basically fixed by the dimension of
space-time and the dimension of the fermion representation. To
get started we may re-scale $\psi$ to get rid of the factor 2 in the fermionic
part of the Dirac-Lagrangian. Moreover, we may introduce a relative
constant $\lambda_{\mbox{\tiny D}}\in\rr$ and thereby replace the total
Lagrangian (\ref{ehymlagrangian}) by
\bb
{\cal L}_{\rm D} &=&
({\bar\psi}i\gamma^\mu\nabla_{\!\!\mu}^{\mbox{\tiny Cl}}\psi +
\lambda_{\mbox{\tiny D}}V_{\rm D})\,\mbox{\small$\sqrt{-|g|}$}\,d^4x.
\ee
Of course, the free parameter $\lambda_{\mbox{\tiny D}}$ can be absorbed by
$\lambda_{\mbox{\tiny gr}},\lambda_{\mbox{\tiny YM}},$ which will then be treated
as arbitrary free parameters. In particular, $\lambda_{\mbox{\tiny gr}}$ will
be proportional to Newton's gravitational constant (or the inverse
square of the Planck length $\ell_{\mbox{\tiny P}}$) after we have introduced
an appropriate length-scale to give the various fields their correct
physical dimensions (see below). Next, we take into account that
all fields are represented in the fermion representation $\rho_{\rm F}.$
Accordingly, there is another numerical constant $\lambda_{\rm F},$ which
only depends on $\rho_{\rm F},$ such that we may re-write the Yang-Mills
part in (\ref{ehympotential}) as
\bb
{\rm tr}{F^{\mbox{\tiny A}}_{\!\mu\nu}}^{\!\dagger}
F_{\!\mbox{\tiny A}}^{\mu\nu}\; =
\;\lambda_{\rm F}
<\!{F^{\mbox{\tiny A}}_{\!\mu\nu}},F_{\!\mbox{\tiny A}}^{\mu\nu}\!>
\ee
Here, $<\cdot,\cdot>$ denotes the Killing form on su(2). Since the
latter is proportional to the usual trace with respect to the
fundamental representation one may re-write again the Yang-Mills
part as
\bb
\lambda_{\mbox{\tiny YM}}\,{\rm tr}{F^{\mbox{\tiny A}}_{\!\mu\nu}}^{\!\dagger}
F_{\!\mbox{\tiny A}}^{\mu\nu} = \frac{1}{2{\rm g}_{\mbox{\tiny YM}}^2}\,
{\rm tr}{F^{\mbox{\tiny A}}_{\!\mu\nu}}F_{\!\mbox{\tiny A}}^{\mu\nu}
\ee
where on the right-hand side $F^{\mbox{\tiny A}}_{\mu\nu}$ is
supposed to be in the fundamental representation of su(2). It follows
that
\bb
\lambda_{\mbox{\tiny YM}} =
\lambda_{\mbox{\tiny rep}}/{\rm g}_{\mbox{\tiny YM}}^2
\ee
with $\lambda_{\mbox{\tiny rep}}$ being a numerical constant which is
basically fixed by the fermion representation and the fundamental representation.
The constant ${\rm g}_{\mbox{\tiny YM}}$ denotes the usual Yang-Mills
coupling constant which parameterizes the most general Killing form on
Lie(G).\\

The specific example discussed so far can be slightly generalized
by taking into account the reducibility  of $\rho_{\rm F}.$
Let $\goth{Z}=\goth{Z}^\dagger$ be the most general element of
the corresponding commutant. In fact, $\goth{Z}$ can be regarded
as a constant mapping commuting with the action of the Clifford
algebra, which can be expressed in terms of the Dirac operator as
follows:
\bb
\label{commutant}
[D,\goth{Z}] = 0.
\ee
Note that, when considered as a condition on the mapping $\goth{Z},$
(\ref{commutant}) cannot be weakened either by the condition
$[D^k,\goth{Z}] = 0$ for some integer $k>1,$  or by $\{D^k,\goth{Z}\} = 0$
for some integer $k\geq 1.$ Indeed, the former case reduces to the condition
(\ref{commutant}) and the anti-commutator condition yields $\goth{Z}=0.$
Basically, the reason is that only the operator $[D,\goth{Z}]$ acts as a
zero order operator on sections $\psi.$\\

Hence, we may generalize the Yang-Mills part in the Dirac-Lagrangian
(\ref{ehymlagrangian}) by
\bb
{\rm tr}(\goth{Z}{F^{\mbox{\tiny A}}_{\!\mu\nu}}^{\!\dagger}
F_{\!\mbox{\tiny A}}^{\mu\nu}).
\ee

The introduction of the commutant with respect to the fermion
representation provides us with a natural parametrization of the
Dirac-Lagrangian which is compatible with the geometrical scheme
of a GTDT. Indeed, the Dirac potential is but a trace of an
endomorphism which is uniquely determined by $D$ (c.f. section 2.2 in
\cite{ToTh:05} and Prop. 3.1 in \cite{Tol:07}). As a result, the Dirac
potential (\ref{diracpotential}) is replaced by
\bb
\label{repdiracpot}
V_{{\rm D},\goth{Z}} &\equiv&
{\rm tr}(\goth{Z}\,r_{\!\mbox{\tiny M}}) + {\rm tr}\!
\left(\goth{Z}\gamma^{\mu\nu}[H_{\mu},H_{\nu}]\right) +
\mbox{\small$\frac{1}{8}$}\,g_{\mu\nu}\,
{\rm tr}\!\left(\goth{Z}\gamma^\sigma\,[H_{\sigma},\gamma^\mu]
\gamma^\lambda\,[H_{\lambda},\gamma^\nu]\right).
\ee
Of course, in the more simple case where $\rho_{\rm F}$ is
irreducible, this replacement just leads back to
$V_{{\rm D},\goth{Z}}=\lambda_{\mbox{\tiny D}}V_{\rm D}$.\\

The parametrization of $D$ depends on its specific form. In
general, one may replace $A_\mu$ by $A'_\mu\equiv\lambda'_{\mbox{\tiny A}} A_\mu$
and (\ref{zeroorderdop}) by
\bb
\label{lambdaphid}
\lambda\Phi_{\mbox{\tiny D}} \equiv\;
\sum_{k=0}^4\;\sum_{0\leq\nu_1<\cdots<\nu_k\leq 3}\,
\gamma^{\nu_1}\cdots\gamma^{\nu_k}\,\lambda_{\mbox{\tiny$(k)$}}
\,\chi^{\mbox{\tiny$(k)$}}_{\nu_1\cdots\nu_k}\;,
\ee with
$\lambda_{\mbox{\tiny$(k)$}}$ being appropriate `` coupling
constants''.\\

For example, in the case of the PDY this corresponds
to the replacement:
\bb
\label{scaledgaugepot}
\theta_{\!\mu} &\mapsto&\theta'_{\!\mu}\equiv
\lambda'\theta_{\!\mu}:=\lambda'_{\mbox{\tiny A}} A_\mu +
\lambda'_{\mbox{\tiny H}}H_\mu,\\[0.1cm]
\label{scaledcurvature}
F^{\mbox{\tiny$\theta$}}_{\!\mu\nu}
&\mapsto& \lambda^{-1}F^{\mbox{\tiny$\theta'$}}_{\!\mu\nu} \equiv
\lambda_{\mbox{\tiny A}}^{-1}F^{\mbox{\tiny A$'$}}_{\!\mu\nu} +
\lambda_{\mbox{\tiny H}}^{-1}F^{\mbox{\tiny A$'$,H$'$}}_{\!\mu\nu}.
\ee
Here, the curvature $F^{\mbox{\tiny$\theta'$}}_{\!\mu\nu}$ is
defined with respect to the parameterized gauge potentials
$A'_\mu$ and $H'_\mu\equiv\lambda'_{\mbox{\tiny H}} H_\mu$. Of course, by
re-scaling the gauge potentials one may assume without loss of
generality that $\lambda'_{\mbox{\tiny A}} = \lambda'_{\mbox{\tiny H}} \equiv 1.$
Therefore, the parameterized PDY corresponds to the replacement:
\bb
\label{parameterizedpauliterm}
F^{\mbox{\tiny$\theta$}}_{\!\mu\nu} &\mapsto&
\lambda^{-1}F^{\mbox{\tiny$\theta$}}_{\!\mu\nu} \equiv
\lambda_{\mbox{\tiny A}}^{-1}F^{\mbox{\tiny A}}_{\!\mu\nu} +
\lambda_{\mbox{\tiny H}}^{-1}F^{\mbox{\tiny A,H}}_{\!\mu\nu}.
\ee

From a geometrical perspective such a parametrization is quite
acceptable, for curvatures are always considered as elements of
vector spaces in contrast to the corresponding gauge potentials.
Moreover, the re-parametrization (\ref{parameterizedpauliterm})
is known from the usual geometrical description of Yang-Mills
gauge theories. However, the constants $\lambda_{\mbox{\tiny A}},
\lambda_{\mbox{\tiny H}}$ should not be identified with the usual
Yang-Mills coupling constants. For example, in the case of
a simple gauge group G and an irreducible fermionic representation
$\rho_{\rm F}$ thereof, the constant $\lambda_{\mbox{\tiny A}}$ turns out
only to be proportional to the Yang-Mills coupling constant
${\rm g}_{\mbox{\tiny YM}}$ which parameterizes the most general
Killing form of G. More precisely, one obtains
\bb
\lambda_{\mbox{\tiny A}} = \sqrt{\frac{\lambda_{\mbox{\tiny YM}}}
{\lambda_{\mbox{\tiny rep}}}}\;{\rm g}_{\mbox{\tiny YM}}
\ee
with $\lambda_{\mbox{\tiny YM}}$ being an arbitrary free parameter.\\

Moreover, as it turns out, one may put $\lambda_{\mbox{\tiny H}}$
equal to one without loss of generality (c.f. (\ref{parameterbeziehung1}) -
(\ref{parameterbeziehung2})). However, the constant $\lambda_{\mbox{\tiny A}}$
will be crucial for the calculation of the Higgs mass (c.f. relations
(\ref{numrel3}) below). Indeed, this freedom will guarantee the
numerical consistence of the presented geometrical description of the STM.\\

Note that the replacement $D\mapsto\lambda D$ is
mathematically inappropriate for $D$ belongs to an affine space.
Moreover, since the relative curvature $F^{\mbox{\tiny A,H}}_{\!\mu\nu}$
decomposes into $F^{\mbox{\tiny A,H}}_{\!\mu\nu}=
F^{\mbox{\tiny H}}_{\!\mu\nu} + \kappa^{\mbox{\tiny A,H}}_{\!\mu\nu},$
one may introduce the more general parametrization:
\bb
\lambda_{\mbox{\tiny H}}^{-1}F^{\mbox{\tiny A,H}}_{\!\mu\nu} \mapsto
\lambda_{\mbox{\tiny H}}^{-1}F^{\mbox{\tiny H}}_{\!\mu\nu} +
\lambda_{\mbox{\tiny int}}^{-1}\kappa^{\mbox{\tiny A,H}}_{\!\mu\nu}.
\ee
However, for reasons of covariance one has to identify
$\lambda_{\mbox{\tiny int}}$ with $\lambda_{\mbox{\tiny H}}.$ Finally,
the parametrization of the off-diagonal elements of (\ref{pdyop})
by the same coupling constants(s) is enforced by quantum field
theory. In fact, it is well-known that the occurrence of
the Pauli-term in the fermionic part of the Lagrangian spoils the
renormalizability of the fermionic theory. It therefore has to drop
out in the fermionic action (c.f. section five in \cite{ToTh:05}).\\

In the {\it parameterized form} the Dirac-Lagrangian
with respect to the above data of the (minimal) STM
explicitly reads:
\bb
\label{parametstmlagrangian}
{\cal L}_{\rm D} &\equiv& {\cal L}_{\rm D, fer} +
{\cal L}_{\rm D, bos}\,,\nonumber\\[0.2cm]
\label{parametfermlagrangian}
{\cal L}_{\rm D,fer} &:=&
({\bar\psi}i\gamma^\mu\nabla_{\!\!\mu}^{\mbox{\tiny Cl}}\psi)\,
\mbox{\small$\sqrt{-|g|}$}\,d^4x +
i({\bar\psi}\gamma_5\phi\psi)
\,\mbox{\small$\sqrt{-|g|}$}\,d^4x,\\[0.2cm]
\label{parametbosolagrangian}
{\cal L}_{\rm D,bos} &:=&
\lambda_{\rm gr}\,r_{\!\mbox{\tiny M}}\,\mbox{\small$\sqrt{-|g|}$}\,d^4x\;
- \;\lambda'_{\mbox{\tiny YM}}\,
{\rm tr}(\goth{Z}{F^{\mbox{\tiny A}}_{\!\mu\nu}}^{\!\dagger}
F_{\!\mbox{\tiny A}}^{\mu\nu})\,\mbox{\small$\sqrt{-|g|}$}\,d^4x\nonumber\\[0.1cm]
&&
+ \;\lambda'_{\mbox{\tiny H}}\,
{\rm tr}(\goth{Z}\nabla_{\!\!\mu}\phi^\dagger\nabla^{\mu}\phi)
\,\mbox{\small$\sqrt{-|g|}$}\,d^4x\nonumber\\[0.1cm]
&&
- \;(\,\alpha'_{\mbox{\tiny H}}\,{\rm tr}\goth{Z}(\phi^\dagger\phi)^2\, -
\,\beta'_{\mbox{\tiny H}}\,{\rm tr}(\goth{Z}\phi^\dagger\phi)\,)
\,\mbox{\small$\sqrt{-|g|}$}\,d^4x\,,
\ee
\bb
\label{parameterbeziehung1}
 \alpha'_{\mbox{\tiny H}}
 &=&
 \frac{27}{64}\frac{1}{\pi \tr (\goth{Z})}
 \left( \frac{\ell}{\ell_{\mbox{\tiny P}}}\right)^2
 \frac{1}{\lambda_{\mbox{\tiny H}}^2} \,,\qquad
 \beta'_{\mbox{\tiny H}}
 =
 \frac{1}{4} \frac{1}{\pi \tr(\goth{Z})}\frac{1}{\ell_{\mbox{\tiny P}}^2}\,,
 \\[0.2cm]
 \label{parameterbeziehung2}
 \lambda'_{\mbox{\tiny YM}}
 &=&
 \frac{1}{8}\frac{1}{\pi \tr(\goth{Z})}
 \left( \frac{\ell}{\ell_{\mbox{\tiny P}}}\right)^2
 \frac{1}{\lam_{\mbox{\tiny A}}^2}\,,\hspace{0.95cm}
 \lambda'_{\mbox{\tiny H}}
 =
 \frac{9}{32}\frac{1}{\pi \tr(\goth{Z})}
 \left( \frac{\ell}{\ell_{\mbox{\tiny P}}}\right)^2
 \frac{1}{\lam_{\mbox{\tiny H}}^2}\,.
\ee

Here, $\psi$ is already appropriately re-scaled and ${\cal L}_{\rm D,bos}$
is normalized such that $\lambda_{\rm gr}=\frac{1}{16\pi\ell_{\mbox{\tiny P}}^2}.$
Note that there are two independent length-scales involved which
are introduced for quite different reasons. First, one arbitrary
length-scale $\ell$ is introduced to provide the various fields involved
with the appropriate physical (length) dimension. As a consequence, one has
then to introduce a second length-scale that is given by the Planck length
$\ell_{\mbox{\tiny P}}$ (Newton's gravitational constant), to make the Einstein-Hilbert
Lagrangian dimensionless. Note that to identify this second length-scale
with the Planck scale is mainly motivated by the ``Newtonian limit'' of Einstein's
theory of gravity. Otherwise, this second length-scale is also considered
as a free parameter. As it turns out, the length-scale $\ell$ corresponds to
the (inverse of the) Higgs mass (please, see below), contrary to what one might
naively expect from (\ref{parameterbeziehung1}). We stress again that all
fields involved are considered to be represented with respect to the fermion
representation $\rho_{\rm F}.$

\section{Higgs Mass Relations}
Since the STM-Lagrangian has a natural ``square root'' in terms of
a PDY, one obtains specific relations between the corresponding
parameters. In the next two sub-sections it will be demonstrated
how these relations yield restrictions to the Higgs mass when
(one-loop) quantum corrections are taken into account.

\subsection{Parameter Relations between GTDT and the STM}
In this sub-section we first re-write the parameterized
Dirac-Lagrangian (\ref{parametstmlagrangian}) in terms of the
ordinary fields of the STM. A comparison with the usual
parametrization of the STM-Lagrangian yields some constraints of
the parameters which are not known in the usual description of the
minimal Standard Model. This discussion is analogous to what has
been presented already in \cite{tol:98}. Hence, we will skip the
details and present here only the relevant results. With the help of
these results, however, we will show how the geometrical description
of the STM in terms of GTDT gives rise to bounds of the Higgs mass.
This, of course, will be discussed in some detail in the next
sub-section.\\

The bosonic Lagrangian within the usual description of the STM reads:
\bb
\label{usualstmlagrangian}
{\cal L}_{\mbox{\tiny SM}} =
&&\left( \frac{1}{2{\rm g}_3^2}\tr (C_{\mu\nu}C^{\mu\nu})
+\frac{1}{2g_2^2}\tr (W_{\mu\nu}W^{\mu\nu})
+\frac{1}{4g_1^2}B_{\mu\nu}B^{\mu\nu}\right)\mbox{\small$\sqrt{-|g|}$}\,d^4 x
\nonumber \\
&& \;+\;
\frac{1}{2}(\na_{\mu}\vp)^* (\na^{\mu}\vp) \mbox{\small$\sqrt{-|g|}$}\,d^4x
- \left[\lam (\vp^*\vp )^2 - \frac{\mu^2}{2} \vp^* \vp \right]
\mbox{\small$\sqrt{-|g|}$}\,d^4x \,.
\label{BosonW}
\ee
Here, respectively, $C_{\mu\nu}$, $W_{\mu\nu}$ and $B_{\mu\nu}$
denote the Yang-Mills field strengths with respect to the fundamental
representation of ${\rm SU(3)}$, ${\rm SU(2)}$ and ${\rm U(1)};$
$\varphi$ is the usual Higgs doublet sitting in the fundamental
representation of ${\rm SU(2)}.$ Its ${\rm U(1)}$ representation
is defined with respect to the hypercharge relations
(\ref{hyperchargerel}), which turn out to be crucial for re-writing
(\ref{parametbosolagrangian}) in terms of the physical fields.\\

In terms of the data of the (minimal) Standard Model the Dirac-Lagrangian of
a PDY (\ref{parametbosolagrangian}) corresponds to the bosonic STM-Lagrangian
(\ref{usualstmlagrangian}) provided the following relations are fulfilled
(see also \cite{tol:98}, \cite{Th:03}):
\bb
\label{gaugecouplingrel}
   \frac{1}{{\rm g}_1^2} = 2A\frac{3Ny_q + y_l \tr X}{4N + \tr X}\,,\qquad
   \frac{1}{{\rm g}_2^2} = A \frac{3N + \tr X}{4N + \tr X}\,, \qquad
   \frac{1}{{\rm g}_3^2} = 4A \frac{N}{4N + \tr X}\,, \label{conI}
  \ee
\bb
\label{conII}
   1 &=& 2B \frac{3 \tr ({\bf g}^{'q}{\bf g}^{'q*} + {\bf g}^q{\bf g}^{q*}) +
         \tr (X{\bf g}^l{\bf g}^{l *})}{4N + \tr X}\,,\\[0.2cm]
\label{treelambdarel}
   \lam &=& C \frac{3 \tr (({\bf g}^{'q}{\bf g}^{'q*})^2 +({\bf g}^q{\bf g}^{q*})^2)
              + \tr(X ({\bf g}^l{\bf g}^{l *})^2)}{4N + \tr X}\,,\\[0.2cm]
   \mu^2 &=& \frac{1}{3\pi} \left( \frac{1}{\ell_{\mbox{\tiny P}}} \right)^2
       \frac{3 \tr ({\bf g}^{'q}{\bf g}^{'q*} + {\bf g}^q{\bf g}^{q*}) +
         \tr (X{\bf g}^l{\bf g}^{l *})}{4N + \tr X}\,.
              \label{conIII}
\ee
Here, we used the following abbreviations:
\bb
y_q := 2(y^q_L)^2 +(y^{d'}_R)^2 +(y^u_R)^2 \,,\qquad
y_l := 2(y^l_L)^2 + (y^l_R )^2,
\ee
\bb
X := \frac{ \lam^l}{\lam_q}\,,
\quad
A := \frac{1}{12\pi}  \left(\frac{\ell}{\ell_{\mbox{\tiny P}}}\right)^2 a^2\,,
\quad
B := \frac{1}{3\pi}  \left( \frac{\ell}{\ell_{\mbox{\tiny P}}}\right)^2 b^2 \,,
\quad
C := \frac{1}{2\pi} \left( \frac{\ell}{\ell_{\mbox{\tiny P}}}
          \right)^2 b^2\,, \label{ABCPARASM}
\ee
with
$a := \frac{1}{\lam_{\mbox{\tiny A}}}$,
$b :=\frac{3}{4}\frac{1}{\lam_{\mbox{\tiny H}}}$ and $\lam_q\in\rr^+$,
as well as $\lam^l:=\diag (\lambda_{1l},\dots ,\lambda_{N l})$ and
$\lambda_{il}\in\rr^+.$  We also made use of
$\goth{Z}={\bf 1}_{4}\ten{\rm diag}(z_L ,z_R),$ with
$z_L :={\rm diag}(\lambda_q {\bf 1}_{\rm 6N},\lambda^l\ten\1_2 )$ and
$z_R :={\rm diag}(\lambda_q{\bf 1}_{\rm 6N},\lambda^l)$ and of the
relations (\ref{hyperchargerel}).\\

From the relations (\ref{conII}) and (\ref{conIII}) one immediately infers that
\bb
\label{higgslengthscale}
\ell_{\mbox{\tiny H}}\equiv\frac{\ell}{\lambda_{\mbox{\tiny H}}} =
2\sqrt{\frac{2}{3}}\,\frac{1}{\mu}\,.
\ee
As mentioned already, we may put $\lambda_{\mbox{\tiny H}}\equiv 1$ for it
simply re-scales $\ell\mapsto\ell_{\mbox{\tiny H}}$ according to the relations
(\ref{parameterbeziehung1}) and (\ref{parameterbeziehung2}).
As a consequence, a PDY is physically parameterized by the two
constants $(\ell_{\mbox{\tiny H}},\lambda_{\mbox{\tiny YMH}})$ instead of
the three parameters $(\ell, \lambda_{\mbox{\tiny A}},\lambda_{\mbox{\tiny H}})$ with
$\lambda_{\mbox{\tiny YMH}}\equiv\lambda_{\mbox{\tiny H}}/\lambda_{\mbox{\tiny A}}.$
Moreover, as far as the Standard Model is concerned, the ``relative
coupling constant'' $\lambda_{\mbox{\tiny YMH}}$ turns out to be
numerically fixed (see below).\\

Because of (\ref{higgslengthscale}), one has
\bb
\ell_{\mbox{\tiny H}}\thicksim 1/{\rm m}_{\mbox{\tiny H}}.
\ee
Hence, the two length-scales involved in the geometrical
description of the (minimal) Standard Model as a specific GTDT are
determined by the Planck mass and the mass of the Higgs. As one
may naively expect, it turns out that
${\rm m}_{\mbox{\tiny H}}/{\rm m}_{\mbox{\tiny P}}\lll 1$ on
one-loop and top-quark-mass approximation. Therefore, within these
approximations $\ell_{\mbox{\tiny H}}$ is the dominant
length-scale and gravitational effects may be fully negligible.\\

The relations (\ref{conI}) -- (\ref{conIII}) are derived on
``tree-level'' by comparing (\ref{parametbosolagrangian}) with
(\ref{usualstmlagrangian}). In this approximation, however, the
constraints for the gauge couplings (\ref{conI}) are inconsistent
with the known experimental data (c.f. Table \ref{Eich}). More
precisely, when taking into account the measured values of the gauge
couplings there exists no choice of the model parameters such that
all three relations for the gauge couplings are fulfilled. On the
other hand, it is well-known that the gauge couplings are running
couplings which depend on the considered energy scale. Hence,
according to the renormalization group philosophy, the inconsistence
of (\ref{conI}) may be interpreted in such a way that (\ref{conI})
-- (\ref{conIII}) are actually supposed to hold true only at certain
critical values of the energy scale. In the following sub-section we
will make use of the renormalization flow equations to determine these
critical energy values. At the critical values it is then possible to
solve the parameter relations with respect to the Higgs self-coupling
$\lam$ from which we eventually obtain the Higgs mass via the ratio
\bb
\label{formhiggs}
\frac{{\rm m}_{\mbox{\tiny H}}}{{\rm m}_{\mbox{\tiny W}}} =
\frac{\sqrt{16\lam}}{{\rm g}_2}\;,
\ee
where the numerical values of the gauge coupling ${\rm g}_2$ and the
mass of the W-boson are regarded to be known from experiments.

\subsection{One-Loop Quantum Corrections and the Higgs Mass}\label{QCHM}
In this section we follow the same strategy as in \cite{CaIoKaSch:97},
\cite{CaIoSch:97} to determine the mass of the Higgs boson.\\

For the STM the renormalization flow equations in one-loop and top-quark mass approximation have been derived in \cite{FoJoStEi:93} using
the $\overline{MS}$-scheme:
\bb
\label{rgei}
\dot{{\rm g}}_1  = \bet_1 ({\rm g}_1 ) := \frac{41}{96 \pi^2}\,{\rm g}_1^3,\quad
\dot{{\rm g}}_2  = \bet_2 ({\rm g}_2 ) := -\frac{19}{96 \pi^2}\,{\rm g}_2^3, \quad
\dot{{\rm g}}_3  = \bet_3 ({\rm g}_3 ) := -\frac{7}{16\pi^2}\,{\rm g}_3^3,
\ee
\bb
\label{rgeq}
\dot{{\rm g}}_t &=& \bet_t ({\rm g}_1 ,{\rm g}_2 ,{\rm g}_3 ,{\rm g}_t )
\;:=\; \frac{1}{16 \pi^2} \left( 9{\rm g}_t^3
-\left(8{\rm g}_3^2 + \frac{9}{4}{\rm g}_2^2 +\frac{17}{12}{\rm g}_1^2 \right)
{\rm g}_t \right),\\[0.2cm]
\label{difflambda}
\dot{\lam} &=& \bet_{\lam}({\rm g}_1 ,{\rm g}_2 ,{\rm g}_3 ,{\rm g}_t ,
\lam )\nonumber\\[0.2cm]
&:=&
\frac{1}{16 \pi^2} \left( 96 \lam^2 +
\left(24 {\rm g}_t^2 -9{\rm g}_2^2 -3{\rm g}_1^2\right)
\lam -6 {\rm g}_t^4 + \frac{9}{32}
{\rm g}_2^4 +\frac{3}{32}{\rm g}_1^4 + \frac{3}{16}{\rm g}_1^2{\rm g}_2^2
\right)\,.
   \ee
Here, the derivative is taken with respect to a dimensionless scale parameter
$t=\ln (\frac{\Lam}{E_0}),$ with $\Lam$ being an arbitrary energy scale and $E_0$
a reference energy. ${\rm g}_t$ is the Yukawa coupling of the top-quark.\\

The renormalization flow equations for ${\rm g}_1$, ${\rm g}_2$, ${\rm g}_3$
can be explicitly integrated:
\bb
\label{eichfunk}
{\rm g}_1(t) = \frac{1}{\sqrt{A_1-\frac{41}{48\pi^2}t}}\,, \qquad
{\rm g}_2(t) = \frac{1}{\sqrt{A_2+\frac{19}{48\pi^2}t}}\,, \qquad
{\rm g}_3(t) = \frac{1}{\sqrt{A_3+\frac{7}{8\pi^2}t}}\,,
\ee
where $A_i\equiv 1/{\rm g}_i^2(0),\quad i=1,\dots,3$.\\

We may use these solutions to determine the critical values $t_c$
for which the relations (\ref{conI}) -- (\ref{conIII}) are
fulfilled. When the relations (\ref{hyperchargeval}) of the
hyper-charges are taken into account one actually obtains a unique
value of the critical energy (N = 3 generations):
\bb
\label{cutofflsg}
t_c &=& \frac{8\pi^2}{21}(3A_1 -9A_2 +4A_3)\\[0.2cm]
&=& 19.32253988 \pm  0.1217255988,
\label{tcval}
\ee
which corresponds to
$E_c = (2.247\pm 0.274 )\cdot 10^{10}\;{\rm GeV}.$ Here, we have again
made use of the experimental values summarized in Table \ref{Eich}.\\

 \begin{table}
  \caption{Gauge Couplings}\label{Eich}
  \begin{center}
  \begin{tabular}{|c|r@{.}l|r@{.}l|} \hline
         & \multicolumn{2}{c|}{value} &
           \multicolumn{2}{c|}{abs. error} \\ \hline
   ${\rm g}_1$ & 0      & 34537  & 0    & 00003 \\
   ${\rm g}_2$ & 0      & 62976  & 0    & 00020  \\
   ${\rm g}_3$ & 1      & 22132   & 0    & 00290  \\  \hline
  \end{tabular}
\end{center}
{\mbox{\small Values of the gauge couplings at the energy
$E_0 ={\rm m}_{\mbox{\tiny Z}} =91.1876\pm 0.0021\;{\rm GeV},$ c.f.
\mbox{\small \cite{Yao:06a}}}}.
  \end{table}

Next, we aim to find an initial value for $\lam$ at the critical scale point $t_c$
and integrate the system (\ref{rgeq}) -- (\ref{difflambda}). This allows to compute
the value of $\lam$ at any scale point $t$ where the values for ${\rm g}_2$ and
${\rm m}_{\mbox{\tiny W}}$ are known.\\

If one divides (\ref{treelambdarel}) and the third equation of
(\ref{conI}) by (\ref{conII}) one obtains in the top-quark
mass approximation:
\bb
\label{numrel3}
\lam &=& \frac{3}{4}{\rm g}_t^2 \,,\qquad
\lam_{\mbox{\tiny YMH}}^2 = \frac{9}{8}\frac{{\rm g}_t^2}{{\rm g}_3^2}\,.
\ee
Note that the parameter $\lam_{\mbox{\tiny YMH}}$ is numerically
fixed by the SU(3) Yang-Mills-coupling constant ${\rm g}_3$ and the
Yukawa-coupling constant ${\rm g}_t$ of the top-quark at the
critical scale point $t_c$ (resp. at the critical energy $E_c$).\\

To proceed, we numerically integrate the system of differential equations
(\ref{rgeq}) -- (\ref{difflambda}). First, we integrate the differential
equation (\ref{rgeq}) for ${\rm g}_t$ with respect to the initial value at $t=0$.
For this, let $t=0$ correspond to the reference energy
$E_0 = {\rm m}_{\mbox{\tiny Z}} = 91.1876\pm 0.0021\;{\rm GeV}.$ We then
calculate the initial values for the top-Yukawa-coupling ${\rm g}_t$ from the
top-mass with help of the relation
\bb
\frac{{\rm g}_t}{{\rm g}_2} =
\frac{1}{2}\,\frac{{\rm m}_{\mbox{\tiny T}}}{{\rm m}_{\mbox{\tiny W}}}
\ee
(\mbox{c.f. \cite{CaIoKaSch:97}}). With the data taken from table (\ref{Eich}) and
${\rm m}_{\mbox{\tiny W}} = 80.403\pm 0.029\;{\rm GeV}$ from \cite{Yao:06c} and
the quark masses from \cite{Yao:06b}:
\bb
 {\rm m}_{\mbox{\tiny T}} &=& 174.2 \pm 3.3 \quad\mbox{GeV}\quad\mbox{(direct observation of top events)},\\
 {\rm m}_{\mbox{\tiny T}} &=& 172.3 \pm 10.2 \quad\mbox{GeV}\quad\mbox{(Standard Model electroweak fit)}
\ee
we obtain as initial values
\bb
{\rm g}_{t}(0) &=& 0.6822142710 \pm 0.01339104334\quad\mbox{(direct observation of top events)},\label{gin1}\\
{\rm g}_t (0)  &=& 0.6747733575 \pm 0.04040821040\quad\mbox{(Standard Model electroweak fit)}.\label{gin2}
\ee
With respect to this solution the value of ${\rm g}_t$ can be computed at the
scale point $t_c$. Next, we numerically integrate (\ref{difflambda}) with respect
to the initial condition:
\bb
 \lam (t_c) = \frac{3}{4}\,{\rm g}_t (t_c)\,.
\ee
Finally, this allows us to calculate $\lam = \lambda(0)$ for the reference
energy $E_0 = {\rm m}_{\mbox{\tiny Z}} = 91.187\;{\rm GeV}$
(c.f. \cite{Yao:06b}):
\bb
\label{aw}
\lam (0) = 0.0661110 \pm 0.0054824 \quad\mbox{(direct observation of top events)},\\
\lam (0) = 0.0647427 \pm 0.0159511 \quad\mbox{(Standard Model electroweak fit)}.
\ee
As a consequence, we obtain for the Higgs mass the values:
\bb
\label{DThval3}
{\rm m}_{\mbox{\tiny H}} &=& 185.6990 \pm 7.6789\;
{\rm GeV}\quad\mbox{(direct observation of top events)},\\
\label{DThval4}
{\rm m}_{\mbox{\tiny H}} &=& 183.7671 \pm 21.4054\;
{\rm GeV}\quad\mbox{(Standard Model electroweak fit)}
\ee
for ${\rm m}_{\mbox{\tiny W}} = 80.403\pm 0.029\,{\rm GeV}$
(c.f. \cite{Yao:06b}, \cite{Yao:06c}). Here, the error is due to the errors of
the initial values of ${\rm g}_1$, ${\rm g}_2$, ${\rm g}_3$, $t_c$ and
${\rm m}_{\mbox{\tiny W}}, {\rm m}_{\mbox{\tiny T}}$ as
well as their influence on the numerical integration of the
renormalization flow equations.\\

According to the STM the Higgs mass can be restricted to the interval
(c.f. \cite{Ro:02}):
\bb
\label{PHB}
{\rm m}_{\mbox{\tiny H}} \in [114 , 193)\;{\rm GeV}.
\ee

It is valid for the STM with one Higgs boson and without super-symmetry. As
a result, the predicted value of the Higgs mass within the geometrical frame
of GTDT is found to be at the upper bound of the interval (\ref{PHB}).\\

\subsection{Model Bounds for the Higgs Mass and Massive Neutrinos}
For the sake of completeness (and comparison, see below), we briefly
discuss here how the statement (\ref{DThval3}) may be weakened if one
introduces the most general parametrization of the PDY. We stress,
however, that such a parametrization is not favored by the
geometrical setup of GTDT (c.f. \cite{ToTh:05}). Disregarding
geometry, however, such a non-geometrical parametrization may be
still of interest for it yields the principal model bounds for the
predicted value of the Higgs mass within the mathematical frame
presented. In this sub-section, we also briefly discuss how the statement
(\ref{DThval3}) may depend on massive neutrinos.\\

From the naive point of view of ``counting free parameters''
one may parameterize (\ref{parameterizedpauliterm}) also as follows:
\bb
\label{genparampauliterm}
\lambda^{-1}\,F^{\mbox{\tiny$\theta$}}_{\!\mu\nu} &\equiv&
\lambda_{\mbox{\tiny A}}^{-1}\,F^{\mbox{\tiny A}}_{\!\mu\nu} + \lambda_{\mbox{\tiny H}}^{-1}\,
(\partial_{\!\mu}H_\nu - \partial_{\!\nu}H_\mu + [A_\mu,H_\nu] -
[A_\nu,H_\mu]) + \lambda^{-1}_{\mbox{\tiny self}}\,[H_\mu,H_\nu]\nonumber\\[0.2cm]
&=&
\lambda_{\mbox{\tiny A}}^{-1}\,F^{\mbox{\tiny A}}_{\!\mu\nu} + \lambda_{\mbox{\tiny H}}^{-1}\,
([\nabla^{\mbox{\tiny A}}_{\!\!\mu},H_\nu] -
[\nabla^{\mbox{\tiny A}}_{\!\!\nu},H_\mu])
+ \lambda^{-1}_{\mbox{\tiny self}}\,[H_\mu,H_\nu]\,.
\ee

It turns out that this non-geometrical parametrization of the PDY does not
change the critical scale-point $t_c.$ In particular, (\ref{genparampauliterm})
does not alter the uniqueness of the critical energy point $E_c$. Nonetheless,
the parametrization (\ref{genparampauliterm}) gives rise to a ``fuzziness'' of
the predicted Higgs mass, similar to what is obtained within the ``(real)
Connes-Lott'' description of the STM (please, see the next section).\\

To obtain the general bounds for the Higgs mass we have to consider the
Dirac-Langrangian for PDY's with the Pauli term being parameterized like
(\ref{genparampauliterm}). One gets the same expressions as in
(\ref{parametfermlagrangian}) and (\ref{parametbosolagrangian}),
however, with different pre-factors. In the case considered they read:
\bb
\label{parameterbeziehung1a}
\alpha'_{\mbox{\tiny H}}
&=&
\frac{27}{64}\frac{1}{\pi \tr (\goth{Z})}
\left( \frac{\ell}{\ell_{\mbox{\tiny P}}}\right)^2
\frac{1}{\lambda_{\mbox{\tiny self}} ^2}\,,\qquad
\beta'_{\mbox{\tiny H}}
=
\frac{1}{4} \frac{1}{\pi \tr(\goth{Z})}\frac{1}{\ell_{\mbox{\tiny P}}^2}\,,
\\[0.2cm]
\lambda'_{\mbox{\tiny YM}}
&=&
\frac{1}{8}\frac{1}{\pi \tr(\goth{Z})}
\left( \frac{\ell}{\ell_{\mbox{\tiny P}}}\right)^2
\frac{1}{\lam_{\mbox{\tiny A}}^2}\,,
\hspace{0.7cm}
\quad\lambda'_{\mbox{\tiny H}}
=
\frac{9}{32}\frac{1}{\pi \tr(\goth{Z})}
\left( \frac{\ell}{\ell_{\mbox{\tiny P}}}\right)^2
\frac{1}{\lam_{\mbox{\tiny H}}^2}\,.
\ee
These relations differ from
(\ref{parameterbeziehung1}) -- (\ref{parameterbeziehung2}) only by the
self-coupling constant $\alpha'_{\rm H}$ due to the parametrization
(\ref{genparampauliterm}).\\

By the same analysis as for the geometrically parameterized PDY one
obtains parameter relations analogous to (\ref{conI}) -- (\ref{conIII}).
They only differ from the latter in the definition of the constants
$A$, $B$, $C:$
\bb
\label{ABCPARASM2}
A := \frac{1}{12\pi}  \left(\frac{\ell_{\mbox{\tiny H}}}
{\ell_{\mbox{\tiny P}}}\right)^2
\lam_{\mbox{\tiny YMH}}^2 \,,\qquad
B := \frac{3}{16\pi}  \left( \frac{\ell_{\mbox{\tiny H}}}
{\ell_{\mbox{\tiny P}}}\right)^2 \,,\qquad
C := \frac{9}{32\pi} \left( \frac{\ell_{\mbox{\tiny H}}}
{\ell_{\mbox{\tiny P}}}\right)^2
\lam_{\mbox{\tiny H,self}}^2
\ee
where again $\ell_{\mbox{\tiny H}}\equiv\ell/\lambda_{\mbox{\tiny H}},
\lam_{\mbox{\tiny YMH}}\equiv\lam_{\mbox{\tiny H}}/{\lam_{\mbox{\tiny A}}}$
and $\lam_{\mbox{\tiny H,self}}:=\lam_{\mbox{\tiny H}}/\lam_{\mbox{\tiny self}}.$\\

Doing the same analysis as in the previous section one concludes that the
value for the critical scale point does not change. Basically, the reason
is that the relations for the gauge couplings (\ref{conI}) remain the same.
For this reason one may proceed in the same way as before to end up with
\bb
\lam_{\mbox{\tiny YMH}}^2 = \frac{9}{8}\frac{{\rm g}_t^2}{{\rm g}_3^2}\,,\qquad
\lam =\frac{3}{4}\lam_{\mbox{\tiny H,self}}^2\; {\rm g}_t^2 \,.
\ee
Therefore, with respect to the more general parametrization
$(\ell_{\mbox{\tiny H}},\lam_{\mbox{\tiny YMH}},\lam_{\mbox{\tiny H,self}})$
the value of Higgs self-coupling constant $\lambda$ at the critical scale point
is not fixed by the appropriate value of ${\rm g}_t(t_c).$ In other words, unlike
to the geometrical parametrization, essentially defined by the length
scale $\ell_{\mbox{\tiny H}}\simeq 1/{\rm m}_{\mbox{\tiny H}},$ the
non-geometrical parametrization gives rise to an additional coupling
constant $\lam_{\mbox{\tiny H,self}},$ that is also tied to the Higgs mass.\\

In what follows we abbreviate
$\kappa := \frac{3}{4}\lam_{\mbox{\tiny H,self}}^2\,{\rm g}_t^2$. We now
have to integrate the flow equation (\ref{difflambda}) with respect to the
initial value:
\bb
\label{BoundDGL}
\lam (t_c) = \kappa\,,\quad \kappa >0 \,.
\ee

The model bounds are then determined by the boundary values for (\ref{rgeq})
with respect to (\ref{BoundDGL}), which give rise to the minimal and
the maximal values for the Higgs mass.\\

Due to standard theorems on ordinary differential equations
the values of the solutions at a certain scale point $t$ of (\ref{difflambda})
depend monotonically on the initial value (c.f. \cite{Am:83}). Hence, the lower
bound for the Higgs mass is determined by integrating (\ref{rgeq}) and
calculating $\underline{{\rm m}_{\mbox{\tiny H}}}$ with respect to
(\ref{formhiggs}) at $\lam (t_c)=0.$ One gets for the considered cases for the top-mass:
\bb
\underline{{\rm m}_{\mbox{\tiny H}}}= 130.952\;
{\rm GeV}\quad\mbox{(direct observation of top events)}\,,\\
\underline{{\rm m}_{\mbox{\tiny H}}}= 113.142\;
{\rm GeV}\quad\mbox{(Standard Model electroweak fit)}\,.
\ee

In order to obtain an upper bound for the Higgs mass one looks for a differential
equation
\bb
\dot{\lam}=\tilde{\betl} \label{upperbound}
\ee
such that:
\begin{enumerate}
\item For the same initial values $\kappa$ the solutions of this equation are
upper bounds of the solutions of (\ref{difflambda}).
\item The solutions explicitly depend on the initial value $\kappa$.
\end{enumerate}
The second property permits to calculate the limit $\kappa\to+\infty$
which yields the upper bound for all possible solutions of (\ref{difflambda}).\\

One may define $\tilde{\betl}$ in the following way:
\bb
\label{upperboundfunction}
\tilde{\betl} &:=& \ell_1\lam^2 + \ell_2 \lam
+ \ell_3\,, \nonumber\\[0.1cm]
\ell_1 &:=& \frac{6}{\pi^2}\,, \qquad
\ell_2 := \frac{1}{16\pi^2} (24\ol{{\rm g}_t}^2 (0)
-9\ol{{\rm g}_2}^2 (t_c +\Del t_c ) -3\ol{{\rm g}_1}^2(0))\,,\nonumber\\[0.1cm]
\ell_3
&:=&
\frac{1}{16\pi^2}(-6\ul{{\rm g}_t}^4 (t_c +\Del t_c )
+\frac{9}{32}\ol{{\rm g}_2}^4 (0)
+\frac{3}{32}\ol{{\rm g}_1}^4 (t_c +\Del t_c ) \nonumber\\[0.1cm]
&&
+\frac{3}{16}\,\ol{{\rm g}_2}^2 (0)\ol{{\rm g}_1}^2 (t_c +\Del t_c ))\,.
\ee
Here, $\Del t_c$ is the error of the critical scale point $t_c$.
We also used the following abbreviations:\\
\bb
\ol{{\rm g}_1}(t) &:=& \frac{1}{\sqrt{A_1 -\Del A_1-\frac{41}{48\pi^2}t}}\,,\qquad
\ol{{\rm g}_2}(t) \;:=\; \frac{1}{\sqrt{A_2 -\Del A_2+\frac{18}{48\pi^2}t}}\,,
\nonumber\\[0.2cm]
\ol{{\rm g}_3}(t) &:=& \frac{1}{\sqrt{A_3 -\Del A_3+\frac{8}{7\pi^2}t}}\,,
\ee
and
\bb
\ul{{\rm g}_1}(t) &:=& \frac{1}{\sqrt{A_1 +\Del A_1-\frac{41}{48\pi^2}t}}\,,\qquad
\ul{{\rm g}_2}(t) \;:=\; \frac{1}{\sqrt{A_2 +\Del A_2+\frac{18}{48\pi^2}t}}\,,
\nonumber\\[0.2cm]
\ul{{\rm g}_3}(t) &:=& \frac{1}{\sqrt{A_3 +\Del A_3+\frac{8}{7\pi^2}t}}\,,
\nonumber\\[0.2cm]
\Del A_i &:=& 2\frac{\Del g_i}{g^3_{i,0}},\quad i=1,\dots ,3\,,
\ee
with
$\Del g_i$ being the error of the initial value $g_{i,0}$ (c.f. table
(\ref{Eich})).\\

Here, $\ol{{\rm g}_t} (t)$ is the solution of the initial value problem
\begin{eqnarray}
\dot{{\rm g}_t} (t) = \frac{1}{16\pi^2}\left(9 {\rm g_t}^3  -
\left(8\ul{{\rm g}_3}^2 +\frac{9}{4}\ul{{\rm g}_2}^2 +
\frac{17}{12}\ul{{\rm g}_1}^2\right){\rm g}_t\right)\,,\quad
 {\rm g}_t = {\rm g}_{t0} +\Del {\rm g}_t\,,
\end{eqnarray}
and $\ul{{\rm g}_t}$ is the solution of the initial value problem
\begin{eqnarray}
 \dot{{\rm g}_t} (t) = \frac{1}{16\pi^2}\left(9 {\rm g_t}^3 -\left(8\ol{{\rm g}_3}^2 +\frac{9}{4}\ol{{\rm g}_2}^2
                           +\frac{17}{12}\ol{{\rm g}_1}^2\right){\rm g}_t\right),\quad
 {\rm g}_t = {\rm g}_{t0} -\Del {\rm g}_t\,,
\end{eqnarray}
with
$\Del {\rm g}_t$ being the error of the initial value ${\rm g}_{t0}$
(c.f. (\ref{gin1}) and (\ref{gin2})).\\

Any solution $\tilde{\lam}$ of (\ref{upperbound}) with initial value
$\kappa >0$ fulfills $\tilde{\lam}(t)\ge\lam (t)$ for $t\ge 0,$ with
$\lam (t)$ being the solution of ({\ref{difflambda}}) with initial
value $\kappa$. Using (\ref{upperboundfunction}) the differential equation
(\ref{upperbound}) can be explicitly solved for arbitrary initial value $\kappa:$
\bb
\tilde{\lam}(t) &=& \sqrt{|\bet|}
\coth \left( \frac{\ell_1\bet}{\sqrt{|\bet|} }(t-t_c )
+ \arcoth \left( \frac{\kappa +\alp}{\sqrt{|\bet|}}\right)\right)
-\alp \label{asloes}\,,\nonumber\\[0.1cm]
\alp &:=& \frac{1}{2} \frac{\ell_1}{\ell_2}\,, \qquad
\bet := -\frac{1}{4} \frac{\ell_2^2}{\ell_1^2} + \frac{\ell_3}{\ell_1}\,,\quad
\bet < 0.
\ee
This permits to calculate
$\tilde{\lam}_{as}(t):=\lim_{\kappa\to\infty}\tilde{\lam}(t)$:
\bb
\tilde{\lam}_{as} (t) &:= \sqrt{|\bet|}
\coth \Big( \frac{\ell_1 \bet}{\sqrt{|\bet|}} \Big.
\Big. (t-t_c ) \Big) - \alp\,,
\ee
which finally yields the upper bound $\overline{{\rm m}_{\mbox{\tiny H}}}$
for the Higgs mass at the scale point $t=0:$
\bb
\tilde{\lam}_{as} (0) &=& 0.463874798\quad \Rightarrow\quad
\ol{{\rm m}_{\mbox{\tiny H}}} = 492.232373 \;
{\rm GeV}\;\mbox{(direct obs. top events)}\,,\\
\tilde{\lam}_{as} (0) &=& 0.459830097\quad \Rightarrow\;
\ol{{\rm m}_{\mbox{\tiny H}}} = 490.081276 \;
{\rm GeV}\quad\mbox{(SM electroweak fit)}\,.
\ee
\phantom{xxx}

Therefore, we end up with the following range of the predicted
value of the Higgs mass:
\bb
\label{higgsmassfuzzrel}
{\rm m}_{\mbox{\tiny H}} &\in& [130.95,\,492.23)\,
{\rm GeV}\quad\mbox{(direct observation of top events)}\,,\\
{\rm m}_{\mbox{\tiny H}} &\in& [113.14,\, 490.08)\,
{\rm GeV}\quad\mbox{(Standard Model electroweak fit)}\,.
\ee

Obviously, this range has a non-empty intersection with (\ref{PHB}).
In particular, it has a lower-bound close to the expected value of
the Higgs mass. Note that the upper bound corresponds to a rough estimate,
only.\\

Next, we discuss a simple modification the STM which takes
into account the possibility of {\it massive neutrinos}.
We restrict ourselves to a few remarks concerning so-called
``Dirac type mass terms'' which seems to fit best with GTDT.\\

The STM can easily be enhanced with a right handed neutrino sector
by replacing $\rho_{F,R}$ in (\ref{frep}) as follows:
\bb
\rho (c,w,\theta ) :=
\diag (c\ten\1_N\ten\diag (e^{iy^{d'}_R\theta},
                              \1_N\ten\diag(e^{iy^e_R
                              \theta},e^{iy^{\nu}_R\theta}))\,,
\ee
with, respectively, $y^e_R ,y^{\nu}_R \in\qq$ being the hyper-charges
of the right handed electron and neutrino. Accordingly, the
matrix $\tilde{\phi}$ in (\ref{higgsfield}) has to be modified by:
 \bb
   \tphi :=\diag
   \left(
   \1_3\ten\left(\begin{array}{cc}
   {\bf g}^{'q}\vp_1 & {\bf g}^q\bp_2 \\
   {\bf g}^{'q} \vp_2 & -{\bf g}^q \bp_1
   \end{array} \right),
   \left( \begin{array}{cc}
   {\bf g}^{'l}\vphi_1 & {\bf g}^{l} \bp_2 \\
   {\bf g}^{'l} \vphi_2 & - {\bf g}^{l} \bp_1
   \end{array}\right)
                \right) ,
  \ee
where ${\bf g}^{'l}$ and ${\bf g}^{l}$ are the corresponding leptonic
Yukawa coupling matrices.\\

This simple modification of the STM permits to construct a theory that
also contains massive neutrinos. The appropriate neutrino mass terms
are generated by so called {\it Dirac type mass terms}. It is known,
however, that there exist other neutrino mass generating mechanisms
as well (c.f. \cite{Bil:02}, \cite{BilGiGrMa:03}).\\

Since Dirac type mass terms result by only modifying the
(right-handed) representation of the gauge group of the STM, one may
perform exactly the same analysis as has been carried out in the
foregoing section. It turns out that the critical energy scale $t_c$
(and hence $E_c$) is identical with (\ref{tcval}). Moreover,
since
the relations (\ref{numrel3}) are unchanged one ends up with the same
predicted value of the Higgs mass (\ref{DThval3}) -- (\ref{DThval4})
as in the STM without massive neutrinos. Note that this is indeed
remarkable, for the parameter relations which correspond to
(\ref{conI}) -- (\ref{conIII}) are nonetheless different from the
relations obtained in the case of massless neutrinos.

%%%%%%%%%%%%%%%%%%%%%%%%%%%%%%%%%%%%%%%%%%%%%%%%%%%%%%%%%%%%%%%%%%%%%%%%%%%%%%%%%%%%%%%%%%%%%%%%%%%%%%%%
\section{A Brief Comparison with NCG}
As mentioned already in the introduction there are various
different geometrical descriptions of the STM. Some of these were
especially addressed to make predictions of the Higgs mass.
Therefore, it may be also of interest to briefly discuss how some
of these approaches to the STM are related to the frame presented
here. In what follows, we will restrict ourselves to two different
geometrical descriptions of the STM within the general frame of
non-commutative geometry. One of which is usually referred to as
``Chamseddine-Connes model (CCM)'' (see, for example
\cite{ChCo:97}, \cite{CaIoKaSch:97}, \cite{CaIoSch:97} and for a version including massive neutrinos \cite{ChCoMa:06}) and which
has some formal similarity to GTDT. The second approach (which can
be actually regarded as the predecessor of CCM) is called the
``Connes-Lott model (CLM)'' (see, for example \cite{CoLo:90},
\cite{SchZy:95}, \cite{KaSch:97}, \cite{IoKaSch:95},
\cite{IoKaSchII:95}, \cite{CaIoSch:99}, \cite{CaIoSch:97},
\cite{IoSch:97}). This model is based on A. Connes' general ideas
of non-commutative geometry as presented, for example, in
\cite{Con:94}, \cite{GV:00} and \cite{SchZy:95}.\\

In the sequel we mainly discuss what is referred to as the ``soft version''
of either of these approaches to the STM. The soft version of CLM and
CCM (in contrast to the so-called ``stiff version'') also takes
into account the possibility of a non-trivial parametrization.
Hence, it is more appropriate to compare these versions of CLM and
CCM with the presented frame of GTDT. Moreover, the
non-parameterized (``stiff'') versions seem physically
inappropriate analogous to ordinary Yang-Mills theory when the
gauge coupling constant is chosen to be equal to one. Such a
non-parameterized version is admissible only with respect to a
purely mathematical discussion of the corresponding geometrical
scheme. Indeed, a non-parametrization usually yields contradictions
with experiments, for it physically corresponds to set (at least some
of) the admissible free parameters of a physical theory equal to one
(see, for instance, \cite{CaIoSch:99}). Of course, a specific geometrical
setup gives sever restrictions to the admissible parametrization of the
scheme considered. Hence, different geometrical descriptions of
the same physical theory (like the STM) may also yield different
constraints on the corresponding parameter set.\\

In this section, the cited values for the Higgs mass almost exclusively
refer to the older assumed value of the top mass of about $175$ GeV. In
order to compare the different NCG approaches with the geometrical
approach proposed in this paper, we mention that in the case of
${\rm m}_{\mbox{\tiny T}}=175\pm 6\;{\rm GeV}$ the predicted value
of the Higgs mass within GTDT reads
\bb
{\rm m}_{\mbox{\tiny H}}=188\pm 15\;{\rm GeV}\,.
\ee
Note that this value also refers to the older values of, respectively,
the W-mass ${\rm m}_{\mbox{\tiny W}} = 80.33\pm 0.15\,{\rm GeV}$ and
${\rm g}_1 = 0.3575,\; {\rm g}_2 = 0.6507,\; {\rm g}_3 = 1.218$ at
$E_0 ={\rm m}_{\mbox{\tiny Z}} =91.187\;{\rm GeV}.$ The appropriate
critical energy scale is given by $E_c = 0.96\cdot 10^{10}\;{\rm GeV}$.

\subsection{Comparison with the CCM}
The formal similarity in the geometrical description of the STM
between CCM and GTDT is that in both approaches Dirac-Yukawa type
operators play a basic role. The motivation, however, is very
different. Within the frame of CCM these generalized Dirac
operators are motivated by non-commutative geometry (via the
tensor product of spectral triples). In contrast, in GTDT the
Dirac-Yukawa type Dirac operators naturally arise from the
Bochner-Lichnerowicz-Weitzenb\"ock decomposition. Physically,
these Dirac operators are motivated by perturbation theory
and the Yukawa coupling. The basic difference of both approaches
lies in the ``action''. Indeed, CCM postulates what is referred
to as {\it spectral action} which incorporates gravity within
non-commutative geometry (c.f. \cite{ChCo:97}). Basically,
the evaluation of the spectral action consists of a
(sophistically) modified heat kernel asymptotic (see, for
instance, in \cite{GV:00}) up to the second non-trivial coefficient
including the ``cosmological constant'' and quadratic Riemannian
curvature terms. As a consequence, the STM Lagrangian is only
reproduced if the base manifold is assumed to be flat and the
cosmological constant is disregarded. Moreover,
for the heat expansion to make mathematically sense one has to deal
with (closed) compact Riemannian manifolds instead of (open)
Lorentzian manifolds to geometrically model ``space-time''
(see also \cite{Sakh:75} and \cite{Sakh:82}).
In contrast, GTDT only uses (globally defined) densities instead
of functionals. Moreover, in the latter scheme a specific
Lagrangian (\ref{basiclagrangian}) is canonically associated
with every Dirac operator (which, in particular, may have
arbitrary signature, c.f. \cite{ToTh:05}). This Lagrangian is
fully determined by the Dirac operator in question.\\

The evaluation of the (soft) spectral action with respect to the
Dirac-Yukawa operator that is defined by
(\ref{fermionrep1}) -- (\ref{hyperchargerel}) leads to
parameter relations which are similar to (\ref{conI}) -- (\ref{conIII}),
see, for example, \cite{CaIoKaSch:97}). In this reference, also
the value of the Higgs mass is calculated by a similar
analysis to that presented in the previous section. It turns
out that the parameter relations for the gauge couplings
(\ref{gaugecouplingrel}) are equivalent to those derived
within CCM. The reason for this is that these relations basically
follow from the fermion representation of the gauge fields. This,
however, is supposed to hold true in both descriptions of the STM.
As a consequence, one obtains the same critical scale point
(\ref{tcval}) and hence also the same critical energy $E_c.$
On the other hand, all other parameter relations turn out to be
essentially different from those presented here. As a
consequence, one obtains a different value for the Higgs mass
(see again \cite{CaIoKaSch:97}, as well as \cite{CaIoSch:99}):
\bb
\label{higgsmassccm}
{\rm m}_{\mbox{\tiny H}} =
190\pm 5\;{\rm GeV}\qquad(\mbox{``soft action''})\,,
\ee
where ${\rm m}_{\mbox{\tiny T}}=175\pm 6\;{\rm GeV}$ and
$E_c = 0.96\cdot 10^{10}\;{\rm GeV}.$\\

Unfortunately, these predictions of the Higgs mass are of limited value
insofar as the CCM approach to the STM is incompatible with certain
experimentally known values of the (ratio of the) gauge and
Yukawa coupling constants. Indeed, the parameter relations of the CCM
(for the so-called ``stiff action'') imply the following relation at
the critical scale-point $t_c$ between the ${\rm SU(3)}-$gauge coupling
constant and the Yukawa coupling constant of the top-quark:
\bb
\label{ccmproblem}
\frac{{\rm g}_3^2}{{\rm g}_t^2} = \frac{3}{2}.
\ee

This relation, however, is not compatible with the known experimental data.
A similar numerical inconsistency is obtained also in the case of the
so-called ``soft-action'' (see again \cite{CaIoKaSch:97}; a detailed
discussion may be found in \cite{Th:03}).  The same holds true
with respect to the value ${\rm m}_{\mbox{\tiny H}}=
175^{\,\mbox{\tiny$+5.8$}}_{\,\mbox{\tiny$-7.8$}}$ (with
${\rm m}_{\mbox{\tiny T}}=178.0\pm 6\;{\rm GeV}$ and
$E_c = 1.1\cdot 10^{17}\;{\rm GeV}$) for the Higgs mass
presented in \cite{KneSch:06}, since this value also refers to the CCM
(c.f. also the discussion about the role of gravity in the conclusion).\\

In \cite{ChCoMa:06} the authors discuss the inclusion of Majorana spinors
in the CCM approach to the STM. From an analysis similar to the one presented
here, the authors arrive at a predicted Higgs mass
\bb
\label{ccmhiggsmass}
{\rm m}_{\mbox{\tiny{H}}} \approx 168 - 170\;{\rm GeV}\,.
\ee
The concrete value depends on the specific assumptions made
on the expected mass of the neutrinos. For example, the value
${\rm m}_{\mbox{\tiny{H}}} \approx 170\;{\rm GeV}$ holds if all but the
top-mass is neglected. The corresponding discussion parallels that
already presented in \cite{KneSch:06}. In contrast, if one of the
neutrino masses is supposed to be of the order of the top-mass,
then ${\rm m}_{\mbox{\tiny{H}}} \approx 168\;{\rm GeV}\,.$\\

However, the value
(\ref{ccmhiggsmass}) of the Higgs mass only follows from the
``stiff'' action of the CCM. As a consequence, the calculated
parameter relations imply the GUT relations between the Yang-Mills
coupling constants:
\bb
\label{gutrelation}
{\rm g}_3^2 = {\rm g}_2^2 ={\mbox{\small$\frac{5}{3}$}}\,{\rm g}_1^2\,.
\ee
As the authors remark, these relations are known to contradict
the measured values of the coupling constants at the W-mass scale.
Indeed, the renormalization flow does not yield the relations
(\ref{gutrelation}) at the GUT energy scale when the measured
values of the gauge coupling constants at the W-mass energy scale
are taken into account as initial conditions. This is usually
interpreted in such a way that there is no ``big desert''. However,
the existence of a big desert is a basic assumption in the CCM and CLM
approach to the STM (like in the geometrical description presented in
this work, c.f. our remarks in the introduction). Actually, in
\cite{ChCoMa:06} the authors do not take into account the second
equality in (\ref{gutrelation}) in order to calculate the value of
the Higgs mass (\ref{ccmhiggsmass}). Indeed, the first equality
in (\ref{gutrelation}) yields a unique critical energy scale of
about $E_c\sim 10^{17}\;{\rm GeV},$ being seven order of magnitudes
higher than the critical energy scale in GTDT. From the corresponding
discussion in \cite{KneSch:06} it follows that the higher the critical
energy scale (i.e. ``the bigger the desert'') the lower the predicted
value of the Higgs mass within NCG.\\

Besides the numerical inconsistence that is implied by the full GUT
relations (\ref{gutrelation}), one may also ask for the
``inner consistence'', for example, with respect to the assumption
that gravitational effects are negligible also on an energy scale
that is only two order of magnitudes below the Planck scale
(c.f. also our corresponding discussion at the end of the paper).
Nonetheless, one may infer from the results presented in
\cite{ChCoMa:06} that adding massive neutrinos to the STM seems
to (slightly) lower the value of the Higgs mass.\\

The inclusion of neutrino masses, as presented in \cite{ChCoMa:06},
may also have the interesting feature to remedy the numerical
inconsistence (\ref{ccmproblem}) in the CCM approach to the STM.
The relation (\ref{ccmproblem}) actually depends on the
assumptions made about the value of the neutrino masses. Hence,
one may ask for the order of magnitude of the neutrino masses that
is necessary to overcome the numerical inconsistence caused by the
relation (\ref{ccmproblem}). Let us call in mind that the latter
relation is derived under the assumption of the validity of the
top-quark mass approximation.\\

In contrast to the relation (\ref{ccmproblem}), in the geometrical frame
presented the above relation between the Yukawa coupling constant and the
${\rm SU(3)}-$gauge coupling constant is replaced by (\ref{numrel3}) which
includes the free ``relative coupling constant'' $\lambda_{\mbox{\tiny YMH}}.$
This additional free para\-meter has its origin in the generalization of the
usual Yang-Mills curvature. It is quite remarkable that the tree-level
relations (\ref{parameterbeziehung1}) -- (\ref{parameterbeziehung2})
are such that only $\lambda_{\mbox{\tiny H}},$ but not $\lambda_{\mbox{\tiny A}},$
can be chosen equal to one without loss of generality. This is because of the
usual Yang-Mills term in the bosonic Lagrangian (\ref{parametbosolagrangian}).
It is this subtle interplay between the usual Yang-Mills curvature and its
generalization with respect to the Higgs gauge potential which allows the
presented geometrical description of the STM to be numerically consistent.

\subsection{Comparison with the CLM}
The Connes-Lott approach to the Standard Model is clearly
conceptually different from the geometrical frame presented here.
The CLM essentially incorporates the basic ideas of A. Connes'
mathematical theory of non-commutative geometry. Hence, it does
not come as a surprise that the appropriate parameter
relations obtained from the CLM are basically different from
those implied by the GTDT approach to the STM
(c.f. \cite{CaIoSch:99}). Yet, in both geometrical schemes
Dirac type operators of the form (\ref{fermionicvacuum})
play a fundamental role though their geometrical origin and
physical interpretation is quite different. Indeed, in the
CLM the geometrical role of the operators (\ref{fermionicvacuum})
is two-fold: First, they correspond to total exterior
derivatives (in this context ${\cal D}$ is referred to as
the ``inner Dirac operator''); Second, (\ref{fermionicvacuum})
induces the non-commutative analogue of the Riemannian volume measure
$\mu_{\mbox{\tiny M}}$ in the bosonic action (``Dixmier trace''). However,
the fermionic and the bosonic CLM-action are defined in totally
different ways, in contrast to the CCM and the frame presented
here. In any case, in the CLM the (Riemannian) metric has to be
chosen by hand, similar to the case of Yang-Mills gauge theories.
Actually, the bosonic action in the CLM frame is a non-commutative
generalization of the usual (Euclidean) Yang-Mills action. Of
course, as far as the calculation of the value of the Higgs
mass is concerned, the metric independence of
(\ref{fermionicvacuum}) does not matter. However, from a purely
conceptual perspective the arbitrariness of the metric seems
unsatisfying (like in the usual Yang-Mills gauge theories)
and may serve as the main motivation to change the definition
of the bosonic action within non-commutative geometry from the
Dixmier trace to the spectral action (c.f. \cite{Con:88},
\cite{Con:96}).\\

During the last decade different CLM approaches to the STM have
been developed. Accordingly, there are also various statements
about the predicted value of the mass of the Higgs within the
frame of non-commutative geometry (see, for example, \cite{Con:95}).
Within Connes' {\it real geometry} the fermionic representation,
considered as an algebra representation, can be chosen differently
from (\ref{fermionrep1}).
As a consequence, the commutant is also differently parameterized.
Moreover, one also obtains more freedom to parameterize the
appropriate scalar products used to define the fermionic and
bosonic actions in the (real) CLM. Note that the introduction of
the real structure also yields a doubling of the gauge degrees
of fermionic freedom quite similar to what is needed in order to
introduce the Pauli-Dirac-Yukawa operator (\ref{pdyop}) (see
also our discussion in \cite{ToTh:05}). Therefore, in contrast to
the CCM and GTDT approach to the STM, within the (real) CLM one
does not obtain a unique critical scale-point $t_c$ on which the
corresponding parameter relations are assumed to hold true but,
instead, a whole range of such points. This range corresponds to
the energy interval
\bb
E_c\in[{\rm m}_{\mbox{\tiny Z}},\;2\cdot 10^5)\;{\rm GeV}.
\ee
Hence, in the (real) CLM the critical energy point $E_c$ is at
least by five orders less than in GTDT (and CCM). Accordingly, the
predicted values of the Higgs mass are contained within the interval
(again, for ${\rm m}_{\mbox{\tiny T}}=175\pm 6\;{\rm GeV}$)
\bb
{\rm m}_{\mbox{\tiny H}}\in(194.5, 291]\,{\rm GeV},\qquad(\mbox{``soft action''})
\ee
which has an empty intersection
with (\ref{PHB}) (c.f. \cite{IoKaSch:95}, \cite{IoKaSchII:95},
\cite{CaIoSch:99}). This, in fact, remains true even if one
restricts the commutant and thus the parametrization in the
CLM. For example, analogous to GTDT there is also a geometrically
distinguished parametrization in (the real) CLM which
gives rise to the following definite value of the Higgs mass
on tree-level (for ${\rm m}_{\mbox{\tiny T}}=175\;{\rm GeV}$ and
$E_c = {\rm m}_{\mbox{\tiny Z}},$ c.f. loc. sit.):
\bb
{\rm m}_{\mbox{\tiny H}} = 289\,{\rm GeV}.
\ee

\section{Conclusion}
In this article, we discussed the possible values of the Higgs
mass as it is predicted by the (minimal) Standard Model when the
latter is considered as a specific gauge theory of Dirac type.
We have shown that this approach to the STM permits
to yield (in a specific approximation including quantum corrections)
a definite value of the Higgs mass without referring to additional
assumptions coming, for instance, from cosmology.
This is quite in contrast to the usual
(non-geometrical) description of the STM, which only gives rise
to a whole range of possible values of the Higgs mass.
Within the GTDT approach to the STM the predicted value of the
Higgs mass is in full accordance with the STM range, though it
lies on the upper bound of the allowed interval. The presented
approach of the STM clearly demonstrates once more the power of
a geometrical understanding of physics and, in the case at hand,
of the Standard Model of particle physics. This is
emphasized by the circumstance, that a non-geometrical
parametrization of a geometrical description may usually yield
a ``fuzziness'' of the predictive power. To demonstrate this we
also discussed the most general (but non-geometrical)
parametrization possible in the GTDT approach to the STM.
Similar to the Connes-Lott approach to the STM this gives rise
to an interval of possible values of the Higgs mass. In the
frame presented, however, this interval has been shown to
have a non-empty intersection with the allowed STM range. In
particular, the GTDT predicted lower bound is close to the
lower bound of the STM range.
In the presented geometrical scheme the fuzziness in the prediction
of the value of the Higgs mass is originated only in the
``self-interaction'' $[H_\mu,H_\nu]$ of the Higgs gauge potential.
Of course, when expressed in terms of the usual Higgs potential
$V_{\!\mbox{\tiny H}}$ it is this self-interaction which gives
rise to the mass of the physical Higgs boson.\\

We also discussed how the predicted value of the Higgs mass may
depend on the existence of massive neutrinos. It turns out that
the inclusion of Dirac type mass terms to the STM does not alter
the results presented. However, the inclusion of Majorana mass
terms in our approach is less straightforward and not taken into
account in this work but will be discussed separately in a
forthcoming paper. Like in \cite{ChCoMa:06} it is expected that
such an inclusion in GTDT will also yield a (slightly) lower
value of the Higgs mass predicted.\\

Since there are some similarities to other geometrical approaches
to the STM, we included a brief comparison of our approach, in
particular, with the Chamseddine-Connes and the (real) Connes-Lott
approach to the STM concerning the Higgs mass.\\

Some comments on the role of gravity within GTDT may be worth
mentioning. Actually, Einstein's theory of gravity is an integral
part of GTDT. Although neglected in our discussion of the Higgs
mass, it plays a fundamental role in this approach to the STM.
This is because it is intimately related to spontaneous symmetry
breaking. Indeed, spontaneous symmetry breaking is considered as
being due to the Higgs gauge potential
$$H_{\!\mu}\thicksim g_{\mu\nu}\gamma^\nu\gamma_5\phi.$$

Accordingly, the role of the
usual Higgs potential $V_{\!\mbox{\tiny H}}$ is regarded as only
giving rise to the mass of the Higgs boson. Concerning
the numerical calculations of the value of the Higgs mass done in
this paper, gravitational effects are assumed to be negligible
(similar to other approaches). The physical reason that this can
be done without contradictions within the setup of GTDT is that
the two length-scales involved, $\ell_{\mbox{\tiny H}}$ and
$\ell_{\mbox{\tiny P}},$ are actually independent of each other.
The drawback of this independence, of course, is that GTDT seems
not to permit a unification of gravity with the strong and the
electroweak interactions of the STM. On the other hand, on the
energy scales considered one may not expect such a unification.
In fact, one obtains for the critical energy point $E_c,$ on
which the parameter relations are shown to hold true, that
$E_c/{\rm m}_{\mbox{\tiny P}}\lll 1.$ Accordingly, one has
${\rm m}_{\mbox{\tiny H}}/{\rm m}_{\mbox{\tiny P}}\lll 1$
(resp. $\ell_{\mbox{\tiny H}}/\ell_{\mbox{\tiny P}}\ggg 1$),
as it is usually expected (and also very much hoped for). At least,
this demonstrates that the GTDT approach to the STM is consistent
with the common assumption that gravity is generically negligible
within the range of validity of the STM, although Einstein's
theory of gravity is naturally included within GTDT. This may
also be inferred from the following rough qualitative considerations.
The Euler-Lagrange equation of the Dirac Lagrangian
(\ref{parametstmlagrangian}) with respect to the metric yields
the Einstein equation with the energy-momentum tensor being
defined by ${\cal L}_{\mbox{\tiny STM}}.$ When all field
excitations are neglected (i.e. putting all fields equal to zero)
this yields the non-vanishing scalar curvature
\bb
\label{scalarcurvvacuum}
r_{\!\mbox{\tiny M}} &=&
-\frac{\pi}{2}\,\frac{(\frac{{\rm m}_{\mbox{\tiny H}}}
{{\rm m}_{\mbox{\tiny P}}})^2}{\lambda}\;
{\rm m}_{\mbox{\tiny H}}^2\nonumber\\[0.2cm]
&\thickapprox& -\frac{10^{-13}}{2\pi}/{\rm cm}^2\,.
\ee

Here, we took into account the values of the Higgs mass
(\ref{DThval3}) and of the Higgs self-coupling constant
(\ref{aw}). Though already small the value
(\ref{scalarcurvvacuum}) should be contrasted with a typical cross
section $\sigma$ of a high energy process such that the
dimensionless product $r_{\!\mbox{\tiny M}}\sigma$ may be
physically interpreted as the quotient of the two relative
accelerations caused by gravitational and high energy effects,
respectively. Roughly, the order of magnitude of a typical
$\sigma$ ranges from $\sigma\thickapprox 10^{-35}\;{\rm cm}^2$
(weak interaction) to $\sigma\thickapprox 10^{-26}\;{\rm cm}^2$
(strong interaction). Hence, \bb |r_{\!\mbox{\tiny
M}}\,\sigma|\lessapprox 10^{-40} \ee which demonstrates again that
gravitational effects are fully negligible on the scale
$\ell_{\mbox{\tiny H}}$. Qualitatively, this does not change even
if field excitations are taken into account, for example, by the
replacement ${\rm m}_{\mbox{\tiny H}}\rightsquigarrow \kappa{\rm
m}_{\mbox{\tiny H}},$ provided that $1\leq\kappa <\kappa_0\equiv
10^{10}.$ Here, $\kappa_0$ is defined such that $|r_{\!\mbox{\tiny
M}}\,\sigma|\thickapprox 1$ where gravitational effects become
comparable with typical high energy effects (of the strong
interaction). Note that the critical energy scale
$E_c\thickapprox 10^{10}\; {\rm GeV}$ corresponds to
$\kappa\thickapprox 10^8.$ Thus, also on the
critical energy scale where the parameter relations
(\ref{gaugecouplingrel}) -- (\ref{conIII}) are assumed to hold
true, the high energy effects still significantly dominate the
gravitational effects since $|r_{\!\mbox{\tiny M}}\,
\sigma|\lessapprox 10^{-8}.$ This, however, does not hold true
any longer for a critical energy scale
$E_c\thickapprox 10^{13} - 10^{17}\;{\rm GeV}$
(c.f. \cite{CaIoSch:99}, \cite{KneSch:06} and \cite{ChCoMa:06}).\\

Irrespective of the concrete value and the geometrical scheme
(``commutative'' or ``non-commutative''),
it seems most remarkable that a prediction of a definite value
of the Higgs mass can be obtained from the pure Standard Model
without additional assumptions, provided the Standard Model is
described in geometrical terms. Of course, that the Standard
Model can be geometrically described at all is certainly in
itself a quite remarkable fact, which one has to take into
account in any theory that aims to go beyond the Standard Model.\\

\begin{appendix}
 \section[Relations to Empirical Parameters]
 {Relations between Gauge Couplings and Empirical Para\-meters}
For the sake of convenience for the reader, in this appendix we call in mind
how the gauge couplings ${\rm g}_1,\,{\rm g}_2,\,{\rm g}_3$ are related to
the experimentally accessible parameters given by the strong coupling constant
$\alpha_{\mbox{\tiny S}},$ the fine structure constant $\alpha$ and the
electroweak mixing angle $\sin^2\vartheta_{\mbox{\tiny W}}.$ For the actual
values of these empirical date we refer to \cite{Yao:06a}.\\

To obtain the relations between the gauge couplings and the above mentioned
empirical parameters we consider the electroweak sector of the fermionic
Lagrangian density of the (minimal) Standard Model (here, the conventions used
are those given in \cite{Na:86}):
\bb
{\cal L}^{\mbox{\tiny int}}_{\mbox{\tiny elw}} =
e\left\{i A_{\mu} J^{\mu}_{\mbox{\tiny em}} +
\frac{i}{\sin \vartheta_{\mbox{\tiny W}}\cos\vartheta_{\mbox{\tiny W}}}
Z_{\mu}J^{\mu}_{\mbox{\tiny NC}} +
\frac{i}{\sqrt{2}\sin\vartheta_{\mbox{\tiny W}}}
(W^{+}_{\mu}J^{\mu}_{\mbox{\tiny CC}} +
W^{-}_{\mu}J^{\mu\dagger}_{\mbox{\tiny CC}})\right\}\,d^4x\,,
\ee
with the currents being defined by:
\bb
J^{\mu}_{\mbox{\tiny em}}
&:=&
\bar{\psi}\gamma^{\mu}(T_3 + Y)\psi\,,\cr
J^{\mu}_{\mbox{\tiny NC}}
&:=&
\bar{\psi}\gamma^{\mu}(T_3 - \sin^2\vartheta_{\mbox{\tiny W}} (T_3 + Y))\psi\,,\cr
J^{\mu}_{\mbox{\tiny CC}}
&:=&
\bar{\psi}\gamma^{\mu}(T_1 + iT_2 )\psi\,.
\ee

As usual, the physical fields are $A_{\mu}$, $Z_{\mu}$ and $W^{\pm}_{\mu},$
where
\bb
A_{\mu} &:=&
\cos\vartheta_{\mbox{\tiny W}}\, W^3_{\mu} -
\sin\vartheta_{\mbox{\tiny W}}\, B_{\mu}\,,\cr
Z_{\mu} &:=&
\sin\vartheta_{\mbox{\tiny W}}\, W^3_{\mu} +
\cos\vartheta_{\mbox{\tiny W}}\, B_{\mu}\,,\cr
W^{\pm}_{\mu} &:=&
\frac{1}{\sqrt{2}}(W^1_{\mu} \mp iW^2_{\mu})\,,
\ee
with, respectively, $W^a_{\mu}$ ($a=1,2,3$) being the gauge fields
of the SU(2) coupling and $T_a$ are the appropriate generators. Here,
$B_\mu$ is the gauge field of the U(1) coupling with generator $Y$
according to the fermionic representation. Note that, in
contrast to the conventions used in the main text of the paper,
the (non-physical) fields $W^a_{\mu}$ and $B_{\mu}$ are re-scaled:
\bb
W^a_{\mu} \to g_2 W^a_{\mu},\qquad B_{\mu} \to g_1 B_{\mu}.
\ee

With these conventions in mind the interaction Lagrangian density of the
electroweak sector of the (minimal) Standard Model reads:
\bb
{\cal L}^{\mbox{\tiny int}}_{\mbox{\tiny elw}} =
\bar{\psi}i\gamma^{\mu}\left(\frac{e}{\cos\vartheta_{\mbox\tiny W}}
B_{\mu}Y + \frac{e}{\sin\vartheta_{\mbox{\tiny W}}}W^a_{\mu}\,T_a
\right)\psi\,d^4x\,.
\ee
Accordingly, the gauge couplings are identified with:
\bb
{\rm g}_1 = \frac{e}{\cos\vartheta_{\mbox{\tiny W}}} =
\frac{e}{\sqrt{1-\sin^2 \vartheta_{\mbox{\tiny W}}}}\,,
\qquad g_2 = \frac{e}{\sin\vartheta_{\mbox{\tiny W}}}\,.
\ee

Finally, using $4\pi\alpha \equiv e^2,$ one gets
\bb
{\rm g}_1 = \sqrt{\frac{4\pi\alpha}{1-\sin^2\vartheta_W}}\,,\qquad
{\rm g}_2 = \sqrt{\frac{4\pi\alpha}{\sin\vartheta_{\mbox{\tiny W}}}}\,.
\ee
Similar to the definition of the fine structure constant, the relation
between the $SU(3)$-coupling ${\rm g}_3$ and the strong coupling constant
$\alpha_{\mbox{\tiny S}}$ reads:
\bb
{\rm g}_3 = \sqrt{4\pi\alpha_{\mbox{\tiny S}}}\,.
\ee
\end{appendix}

\vspace{1.5cm}

\end{document}